\documentclass[12pt,a4paper]{article}

\usepackage[british]{babel}

\usepackage[a4paper,top=2cm,bottom=2cm,left=2.5cm,right=2.5cm,marginparwidth=1.75cm]{geometry}

\usepackage[style=apa, backend=biber]{biblatex} 
\DeclareLanguageMapping{british}{british-apa} 
\DeclareFieldFormat[article]{volume}{\apanum{#1}} 

\usepackage{amsmath}
\usepackage{graphicx}
\usepackage[colorlinks=true, allcolors=blue]{hyperref}
\usepackage{hyperref}
\usepackage[title]{appendix}
\usepackage{mathrsfs}
\usepackage{amsfonts}
\usepackage{booktabs} 
\usepackage{caption}  
\usepackage{threeparttable} 
\usepackage{algorithm}
\usepackage{algorithmicx}
\usepackage{algpseudocode}
\usepackage{listings}
\usepackage{enumitem}
\usepackage{chngcntr}
\usepackage{booktabs}
\usepackage{lipsum}
\usepackage{subcaption}
\usepackage{authblk}
\usepackage[T1]{fontenc}    
\usepackage{csquotes}       
\usepackage{diagbox}
\usepackage{tabularx}
\usepackage{physics}

\usepackage{setspace}
\usepackage{titlesec}
\titleformat{\section} 
  {\normalfont\Large\bfseries}{\thesection.}{1em}{}
  
\usepackage{float}   
\usepackage{caption} 


\title{Characterizing the Inter-Core Qubit Traffic in Large-Scale Quantum Modular Architectures}

\author[1,*]{Sahar Ben Rached}
\author[1]{Isaac Lopez Agudo}
\author[1]{Santiago Rodrigo Muñoz}
\author[2]{Medina Bandic}
\author[4,5]{Artur Garcia-Saez}
\author[2]{Sebastian Feld}
\author[2]{Hans van Someren}
\author[1]{Eduard Alarc\'on}
\author[3]{Carmen G. Almudéver}
\author[1]{Sergi Abadal}
\affil[1]{\small Universitat Polit\`ecnica de Catalunya, Carrer de Jordi Girona, 31, 08034, Barcelona, Spain}
\affil[2]{Delft University of Technology, Mekelweg 5, Delft, South Holland, The Netherlands}
\affil[3]{Universitat Polit\`ecnica de Val\`encia, Camí de Vera, s/n, 46022, Val\`encia, Spain}
\affil[4]{Barcelona Supercomputing Center, Plaça Eusebi Güell, 1-3, 08034 Barcelona, Spain}
\affil[5]{Qilimanjaro Quantum Tech., Carrer dels Comtes de Bell-Lloc, 161, 08014 Barcelona, Spain}
\affil[*]{Corresponding author: \texttt{sahar.benrached@upc.edu}}

\date{}  

\begin{document}
\maketitle

\begin{abstract}
Modular quantum processor architectures are envisioned as a promising solution for the scalability of quantum computing systems beyond the Noisy Intermediate Scale Quantum (NISQ) era. Based upon interconnecting tens to hundreds of quantum processors (i.e cores) via quantum coherent and classical links, this approach unravels the pressing limitations of densely qubit-packed monolithic processors, mainly by mitigating the requirements of qubit control and enhancing qubit isolation. Therefore, this new architectural design enables executing large-scale algorithms in a distributed manner. In order to assess the performance and optimize such architectures, it is crucial to analyze the quantum state transfers occurring via inter-core communication networks, referred to as inter-core qubit traffic. This would provide insights to improve the software and hardware stack for multi-core quantum computers. To this aim, we present a characterization of the spatio-temporal inter-core qubit traffic for different large-scale quantum algorithms. The programs are compiled on an all-to-all connected multi-core architecture that supports up to around 1000 qubits. We characterize the qubit traffic based on multiple performance metrics to assess the computational process and the communication overhead. Based on the showcased results, we conclude on the parallelization and scalability of presented algorithms, qualitatively evaluate the resource requirements as we scale circuit sizes, and lay the foundations of application-oriented benchmarking of large-scale multi-core architectures. 
\end{abstract}



\section{Introduction}

\begin{figure*}[t!]
  \centering
  \includegraphics[scale=0.45]{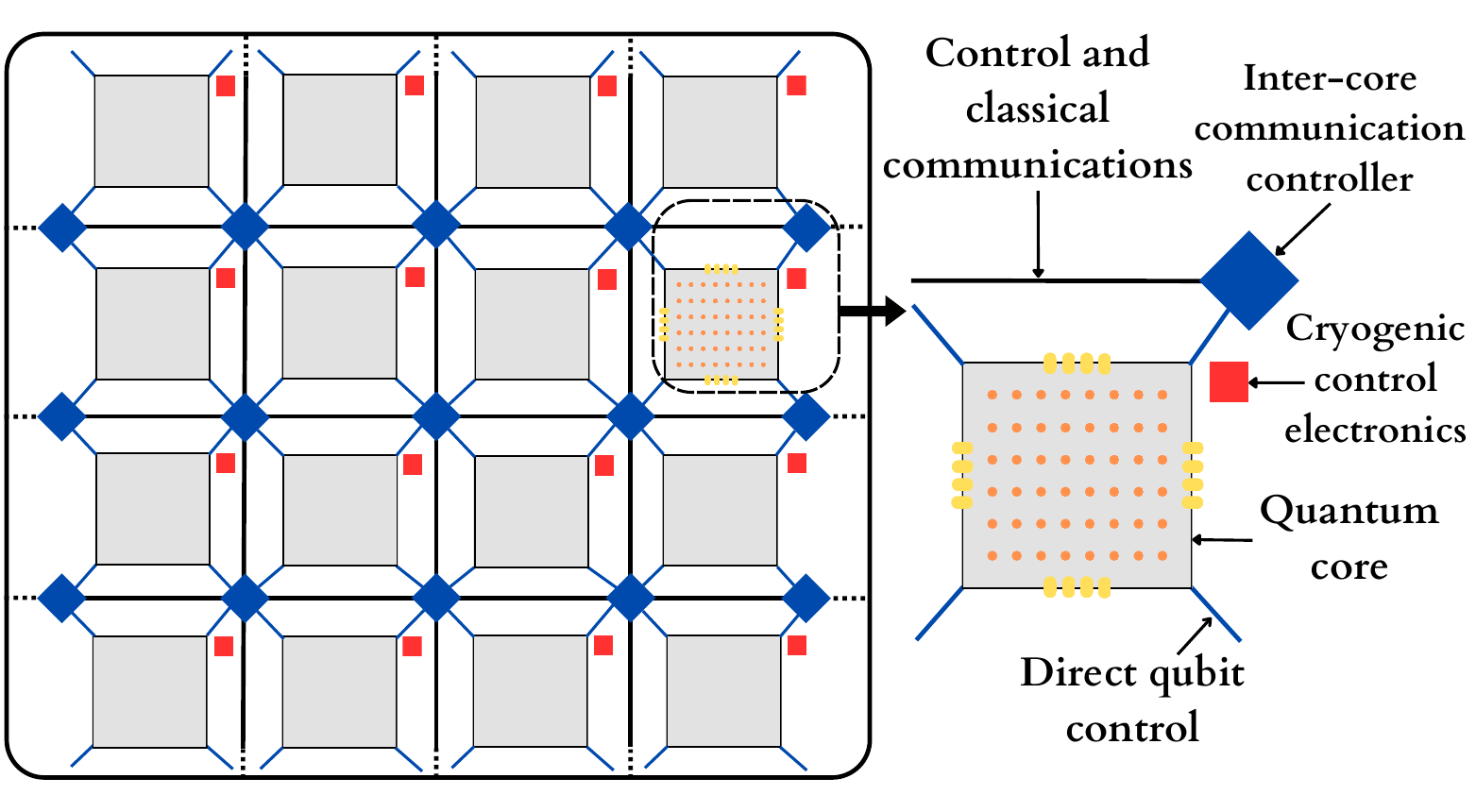}
  \caption{Overview of the envisioned modular quantum computer architecture.}
  \label{mc-architecture}
\end{figure*}

Quantum computing technology has witnessed a substantial progress in hardware development and software design in pursuit of fault-tolerant quantum computers (\cite{b1}). In principle, unlocking their long-awaited computational power is contingent on integrating thousands to millions of qubits into robust processors. Yet, densely-packed monolithic quantum processors raise several issues that deteriorate the computational results; mainly due to the effect of crosstalk, quantum state disturbance, and increased complexity of the control system (\cite{b2}). Building multi-core quantum processors has been introduced among the most promising strategies to scale the contemporary Noisy Intermediate Scale Quantum (NISQ) devices (\cite{b3}) considering the various types of processor technologies (\cite{b4, b5, b6, b7, b8, b9}). The main aim of this approach is addressing the aforementioned issues and scaling quantum processors to a size where complex problems could be practically solved.

In quantum computing, multi-core processors are designed by interconnecting smaller-sized cores with tens to hundreds of qubits via a chip-scale intranet that enables the implementation of remote gates and quantum state transfer between cores (\cite{b10}), as illustrated in Figure \ref{mc-architecture}. This technique presumably alleviates the qubit control requirements and enhances qubit isolation. Therefore, modular architectures are envisioned to facilitate the integration of thousands to millions of qubits in a single system.

However, building multi-core processors still comes with its own set of challenges as quantum communications are highly latency-prone and impose additional technical inefficiencies to designing processors (\cite{b11}). Indeed, integrating chip-scale networks imposes a new paradigm as we need communication channels that maintain the quantum information and the properties of superposition and entanglement, rendering classical communication technologies impractical. Additionally, quantum state retransmission is not possible due to the fundamental limitation that forbids copying data due to the inherent qubit properties as stated by the no-cloning theorem (\cite{b12}). Most critically, communication latencies induce a higher risk of data losses during the qubit transmission operation among cores due to state decoherence. Several methods have been proposed to build inter-core communication networks for modular quantum computers considering the aforementioned limitations and the diversity of technology platforms, such as integrating quantum links to interconnect superconducting chips (\cite{b13}), ion-shuttling for ion-trapped computers (\cite{b14}), photonic networks (\cite{b15}) and the teleportation protocol setup for state transfer (\cite{b16}). Besides, the requirements for modularity extend beyond the qubit layer to impose further constraints at the networking, control, and compilation layers, hence the necessity of a consolidated software-hardware stack (\cite{b17}). 

The multi-core quantum approach is termed after the technique traditionally used in classical computing to enhance the performance of CPUs owing to multitasking (\cite{proceedings}), energy efficiency (\cite{b19}), on-demand hardware scalability, improved resource utilization, and parallel processing for specific tasks (\cite{b20}). In classical multi-core computing, designing efficient Networks-on-Chip (NoCs) and inter-core networking is of significant importance as it directly impacts the overall processor performance. To build a network properly, it is critical to understand the traffic it serves. Consequently, substantial efforts have been dedicated to characterizing the communication occurring within multi-core systems and the adapted applications. Early research by Soteriou et al. (\cite{b21}) focused on analyzing a range of multiprocessors with 16 to 32 cores, using standard benchmark suites such as SPEC or PARSEC to explore temporal burstiness, spatial hotspotness, and source-destination distance. Barrow et al. (\cite{b22}) relied on the same methods to analyze the memory sharing patterns leading to certain traffic characteristics. Later studies extended the analyses to larger systems with up to 64 cores, examining specific aspects like time-varying traffic characteristics (\cite{b23}), in addition to the impact of specific architectural choices such as the cache coherence protocol on the communications (\cite{b24,b25}). The workload characterization studies notably advanced the chip-scale networks field. Indeed, such contributions allowed researchers to analyze the correlation between particular traffic characteristics and on-chip network congestion (\cite{b26}) or to design better NoCs, topologies, and routing mechanisms at the chip scale. A similar characterization of the traffic is lacking in the quantum computing domain. 

As the early prototypes of quantum modular architectures are emerging, it is pertinent to move beyond existing characterization efforts for monolithic processors and to employ analogous characterization techniques to classical computing while accounting for the fundamental differences between classical and quantum multiprocessors. Characterizing the performance of quantum computers has been advancing as the technology is developing, focusing mainly on benchmarking monolithic processors as will be detailed in Section \ref{benchmarking}. For instance, the work in (\cite{b27}) proposed methods for profiling quantum circuits based on their interaction graph parameters, aiming to improve compilation techniques. While the algorithms were executed on monolithic processors hosting a limited number of qubits, this work gives insight into the circuit properties to consider for evaluating compilation algorithms and quantum processors. Assessing the system performance requires defining a set of relevant and representative metrics. In (\cite{b28}), a set of performance metrics to characterize the spatio-temporal qubit routing in connectivity-constrained monolithic architectures is proposed, and in modular architectures at a limited scale as well (\cite{b29}). In the latter, a set of tools to extract and analyze the inter-core qubit traffic of quantum algorithms executed on multi-core processors of 128 qubits using OpenQL compiler is introduced.

In this paper, we improve the methodology from our prior works by increasing the number of qubits, executing more application-oriented algorithms, extending the set of performance metrics to specifically assess the inter-core qubit communication resources, and analyzing the scalability of modular architectures in the strong and weak scaling regimes. On this account, we present a pioneering inter-core qubit traffic analysis in multi-core architectures hosting around 1000 qubits by executing quantum algorithms of various structures. We interpret the inter-core qubit traffic as the qubit movement among cores orchestrated by the compiler to execute programs on modular architectures considering a given circuit mapping algorithm and a particular communication protocol. We conduct a thorough analysis of each compiled circuit's computation and communication workloads according to performance metrics of resource usage uniformity and spatio-temporal locality. We select our performance metrics based on current benchmarking techniques in quantum computing, and add metrics that specifically represent inter-core communications. Therefore, our main contributions include introducing performance metrics for resource assessment and system scalability analysis, and setting the foundations of an application-oriented benchmarking of large-scale modular quantum processors.

This paper is organized as follows: in Section \ref{related-work}, we start by presenting the on-going research on methods for assessing quantum systems and developing compilation techniques for modular quantum computers. In Section \ref{methodology}, we propose our work methodology, the guidelines to define practical and scalable performance metrics for characterizing modular quantum processors, present our selected algorithms covering diverse applications, and introduce our set of performance metrics. The remainder of the paper showcases the results for circuit mapping and qubit traffic analysis in Section \ref{results}, followed by a thorough discussion in Section \ref{discussion} on the impact of communication overhead on the overall computational process and the scalability of quantum algorithms in modular architectures. We close with the main conclusions and future work in Section \ref{conclusion}.
\section{Related work} \label{related-work}

\subsection{Benchmarking quantum computers} \label{benchmarking}
Benchmarking quantum processors has been evolving with the continuous progress of the field, focusing solely on monolithic architectures. In (\cite{b30}), guidelines are proposed to standardize benchmarks, ensuring they are randomized, well-defined, holistic, and device-independent. We discern various proposals for benchmarking frameworks covering the different layers of the system from the low-level qubit layer to high level applications, taking into account the diversity of qubit technologies: 

\vspace{0.1cm}
\noindent \textbf{Physical benchmarking:}
Considering the high risk of quantum state loss due to imperfect control and qubit interactions, evaluating the reliability of the computational processes in the presence of noise is necessary to determine the efficiency of a given device. There exists typical parameters that reflect the quality of the qubits and the overall performance of the control system, mainly the relaxation time T1, the dephasing time T2, single-qubit gate fidelity, two-qubit gate fidelity, and readout fidelity. These parameters are used in the quantum process tomography technique (\cite{b31}) to characterize the performance of any computation process, as well as randomized benchmarking (\cite{b32}), which is an experimental method for measuring the average error rates of quantum computers by implementing long sequences of randomly sampled quantum gate operations. However, both of these methods scale poorly with the number of qubits and are only effective to evaluate processors with a small number of qubits. PyGSTi (\cite{b33}) is also a toolkit designed for evaluating and characterizing the performance of quantum computing processors, specifically qubits, errors and drift, by executing quantum characterization, verification, and validation (QCVV) protocols. While physical benchmarks represent the properties of the device and are essential to detect points of improvement, they overlook the performance of the processor in solving algorithms and executing circuits.

\vspace{0.1cm}
\noindent \textbf{Synthetic benchmarking:}
The efficiency of quantum computers is dependent on the performance of the system in executing circuits, especially for prototype devices. There are currently standard benchmarks for evaluating the NISQ devices' computational performance, such as the CLOPS (\cite{b34}) benchmark to evaluate speed and the Quantum Volume (\cite{b35}), which is a single metric that considers several parameters from each layer of the system to measure its quality as indicated by the largest square random quantum circuit that can be executed reliably. A generalized framework of the Quantum Volume is referred to as Volumetric Benchmarks (VBs) (\cite{b36}), which execute rectangular circuits. Yet, the reliance of these metrics on randomness and physical properties of the system limits their significance and scalability. A similar assessment we find particularly for ion-trapped computers is the Algorithmic Qubits (\cite{b37}). It uses quantum algorithms rather than random circuits to determine the largest number of effectively perfect qubits that can be used for a program, considering error-correction and the qubit count. 

\vspace{0.1cm}
\noindent \textbf{Application-oriented benchmarking:}
Application-level benchmarks rely on the program results accuracy as an indicator of the quantum computer's computational capabilities and scalability. Due to the diversity of applications and their purposeful usage, this metric stands as a fundamental aspect for designing several widely-used benchmarking sets. An application-oriented benchmarking suite is proposed in (\cite{b38}) aiming to estimate the output fidelity, quality of and the time to solution. The majority of such benchmarks use Variational Quantum Circuits (VQCs) to evaluate certain properties of the system, such as QPack (\cite{b39}) that implements the Max-Cut, dominating set, and Travelling Salesman Problem (TSP) circuits to capture runtime, best approximation error, success probability, and scalability. Additionally, QASMBench (\cite{b40}) is a cross-platform suite that executes circuits of different problems to determine various circuit properties such as gate distribution, parallelism, and communication overhead. SupermarQ (\cite{b41}) is also a benchmarking suite that includes diverse applications and evaluates real quantum devices of different platforms according to a customized set of performance metrics such as program communication, critical depth, and liveness. The quantum LINPACK benchmark (\cite{b42}) is developed to measure the performance by its capacity to solve random systems of linear equations with dense random matrices, which may not capture the ability of the device to solve various applications. 

The aforementioned benchmarks are applicable exclusively for monolithic QPUs and lack metrics that assess the communication workload in multi-core systems. For our work, they are useful in perceiving metrics that assess the performance of the system. We extend the current performance assessment metrics to include inter-core communication cost and overhead, not merely computation resources.

\subsection{Compilation techniques for modular architectures} 
Compilers play a critical role in the quantum computing stack as they translate high-level quantum programming languages into executable instructions on qubits and apply the necessary modifications to the input circuit in order to meet the underlying hardware constraints. Due to fundamental differences between quantum and classical computation, quantum compilers are designed to comply with the circuit-based quantum algorithm implementation and the problem size, and to account for the hardware technology properties, mainly qubit control and readout techniques and the processor topology.

Quantum compilers typically support specific programming languages which provide an abstract representation of circuit gates, such as Qiskit (\cite{b43}), Cirq (\cite{b44}), Q\# (\cite{b45}), among others. Compilers are mainly responsible of decomposing the input operations to native gates supported by the hardware device. In the process of translating operations to their low-level representations, compilers apply gate optimization techniques aiming to improve the efficiency and performance. Quantum circuit mapping is also an important aspect of compilers, especially that contemporary hardware architectures often have limited connectivity between qubits. This aims to rearrange qubit placement to allow the direct application of native two-qubit gates - mainly CNOT gates. Initial placement is usually employed to map virtual qubits from the high-level program to physical qubits on the device, taking into account the underlying connectivity constraints. Additionally, compilers handle qubit routing and scheduling for executing gate operations optimally. For future fault-tolerant quantum computers, compilation techniques will include quantum error correction and fault tolerant protocols to reduce the impact of errors and improve the overall reliability of quantum computations.

Designing compilers for multi-core architectures specifically takes into account the inter-core operations necessary to apply two-qubit gates on qubits initially placed in different cores, for example by applying remote two-qubit gates (\cite{b46}) or inter-core quantum state transfer to bring interacting qubits into physical proximity (\cite{b47}). Compilation algorithms for multi-core processors are coming into play to accommodate the various architectural constraints and communication requirements as the technology is developing. Several methods are under investigation, such as the approach presented by Baker et al. in (\cite{b48}) aiming to minimize non-local communication overhead by defining optimized mappings of virtual qubits to physical ones for each timeslice, taking into account the cost of moving data between timeslices and using a tunable lookahead scheme to anticipate and reduce future data movement costs. Other proposals include quantum circuit cutting and reconstruction (\cite{b49}), QUBO-formulated circuit mapping optimization approach (\cite{b50}), and qubit-to-core assignment considering the interactions between qubits over the circuit based on the Hungarian algorithm (\cite{b51}). The communication resources should be characterized and optimized at the compilation level, since the inter-core communication technique and scheduling determines the overhead and directly impacts the computational process. 
\section{Methodology} \label{methodology}

\subsection{Work methodology}

\begin{figure*}[t!]
  \centering
  \includegraphics[scale=0.3]{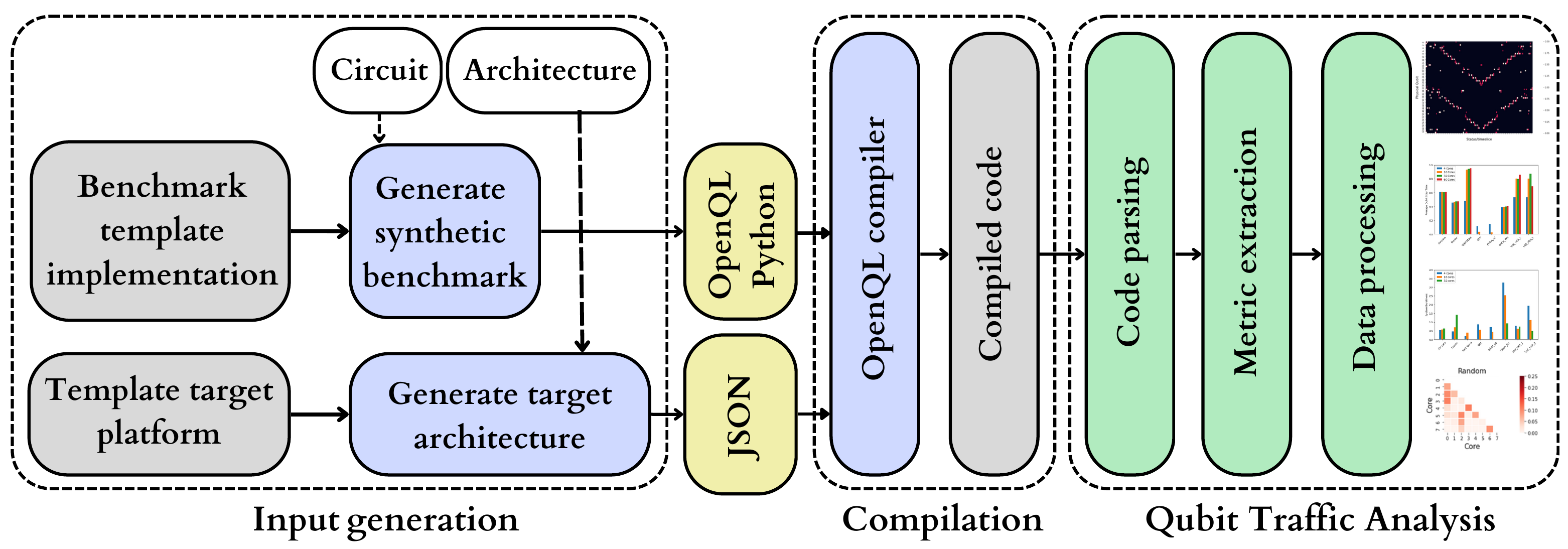}
  \caption{Flow diagram of the qubit traffic analysis software tool.}
  \label{methodology-diagram}
\end{figure*}

The primary goal of this study is to analyze the inter-core qubit traffic in modular quantum processors to assess their scalability. To achieve this, we follow the methodology depicted in the Figure \ref{methodology-diagram}. We compile different quantum circuits on various architecture sizes to generate inter-core communication traces. These traces are then analyzed to extract relevant data that formulate our performance metrics, which are further processed and graphically visualized.

For our work, we use OpenQL (\cite{b52}) to compile and optimize quantum circuits. The inter-core state transfer protocol applied is the teleportation-swap. This consists of applying two simultaneous teleportation operations in opposite directions between two cores to move a computation qubit into proximity with a second qubit in a different core and enabling two-qubit gate application. The mapping algorithm embedded in OpenQL is proposed by Baker et al. in (\cite{b48}). This mapping algorithm assumes chips of tightly connected qubits, with sparser connections between these chips. In this work, we consider a device architecture supporting an intra- and inter-core all-to-all connectivity. In this case, optimal initial placement and qubit routing are not needed, and the compiler initially maps virtual qubits directly to physical qubits. This simplifying assumption is primarily imposed by the compilation algorithm employed. Hence, we provide a theoretical lower bound of the performance of inter-core communication that could be used as a standard for evaluating the performance of more practical architectures which are constrained by qubit and core topologies. We also assume any idling qubit which state is not used for computation operations can be utilized as a communication qubit, and therefore favoring simultaneous application of inter-core communication operations.

In analogy to the scalability analysis of classical multi-core computers, we implement circuits on modular architectures in two scaling regimes:
\begin{itemize}
    \item \textbf{Strong scaling:} we fix the number of qubits per core and increase the number of cores at the same time that we scale the application to fit within the increasing number of qubits. This aims to evaluate the communication overhead as we add more cores, and its impact on the computation process. We fix the size of a single core to 16 qubits, and increase the number of cores and the size of circuits until around 1000 qubits.
    \item \textbf{Weak scaling:} we keep the same problem size, i.e total number of qubits, and increase the number of cores. We inspect the performance of the system in the weak scaling to assess how the program performance varies with the number of cores for a fixed problem size. In this case, the number of qubits per core varies, yet the total number of qubits is counting 512.
\end{itemize}
We present the results of each circuit up to the largest size we could successfully compile.

\subsection{Benchmark Selection Criteria} \label{criteria}
For the purpose of analyzing inter-core qubit traffic and determining the communication overhead, we evaluate the execution performance according to a customized set of metrics that considers both computation and communication resources as explained in Section \ref{metrics}.

We base our selection of applications upon four primary design guidelines:

\begin{itemize} 
\item \textbf{Diversity of circuit structure}:
different algorithms have different requirements and resources for an optimal execution on a modular architecture. Based on our analysis, we aim to correlate the algorithm structure to the resource requirements. We therefore select a wide range of circuits, from structured to random algorithms, in order to analyze their operational constraints and communication overhead.
\item \textbf{Application-oriented}: 
The multi-core processors approach aims to scale quantum computers to the size where they support large-scale real-world applications. We therefore select state-of-the-art sub-routines and algorithms envisioned to solve complex problems in physics (\cite{b53}), chemistry (\cite{b54}), finance (\cite{b55}), machine learning (\cite{b56}), and database search (\cite{b57}), among others.
\item \textbf{Scalability}:
The standard benchmarking techniques and performance metrics, such as Quantum Volume and randomized benchmarking, are not adequate to probe large-scale modular devices as their performance becomes intractable with the increasing number of qubits. Therefore, an application-oriented benchmark analyzed with a representative set of metrics is presumably applicable in large-scale systems.

\item \textbf{Inter-core communication density}:
We select algorithms with a high density of two-qubit gates and qubit operations applied across cores. We analyze the communication overhead originating from the inter-core state exchange.
\end{itemize}

\subsection{Selected benchmarks}
Quantum algorithms are better suited for evaluating the overall performance of the system, from the software to the hardware stack, than circuit or gate-based methods which rather assess the hardware performance. Additionally, at the time of writing this paper, no modular quantum computing system hosting 1,000 qubits is available, rendering hardware evaluation methods not applicable. The selected circuits will be useful for a wide range of applications and domains where quantum computing is expected to be advantageous. We present the circuit of each algorithm implemented in Appendix \ref{implemeted-circuits}.

We summarize the structure of each circuit implemented in Table \ref{structures}, showcasing the two-qubit interconnectivity, i.e how the two-qubit gates are applied in the circuit, and the estimated gate count when scaling them up, referring to the upper bound on the number of gates. Further details on the gate count estimation is found in Appendix \ref{gate-count}. We inspect the circuit structure to analyze the evolution of computation and communication resources as we increase the program size.

\begin{table*}
    \centering
    \begin{tabular}{ccc}
    \hline
        \textbf{Circuit} & \textbf{Two-qubit interconnectivity} & \textbf{Estimated gate count}\\
    \hline
        \textbf{Cuccaro Adder} & Nearest-neighbor & $\approx O(N)$\\
        \textbf{Grover} & $\frac{N}{2}$ to ($\frac{N}{2}+1$) qubits & $\approx O(k \times N)$ \\
        \textbf{GHZ State} & One qubit to all & $\approx O(N)$\\
        \textbf{QFT} & ($N-q$) qubits to one & $\approx O(N^{2})$\\
        \textbf{QAOA MaxCut Ansatz} & Nearest-neighbor & $\approx O(l \times N)$\\
        \textbf{VQE HEA\_1 and HEA\_2} & Nearest-neighbor & $\approx O(l \times N)$\\
    \hline
    \end{tabular}
    \caption{Estimated circuit structures of the implemented applications. $N$, $q$, $k$, and $l$ indicate the total number of qubits, qubit index, number of iterations in the Grover's circuit, and number of ansatz layers, respectively. The exact gate count depends on the circuit implementation, which may involve additional gates for initialization, measurement, and optimization techniques introduced by the compiler.}
    \label{structures}
\end{table*}

\subsubsection{Cuccaro adder}
The Cuccaro adder is an arithmetic operator designed for adding binary numbers of equal bit-string size using quantum computers (\cite{b58}). The algorithm is based on the ripple-carry approach built using one ancilla qubit and by applying sequentially dependent CNOT and CCNOT gates, forming the Majority (MAJ) gate that computes the majority of three bits in place and the “UnMajority and Add” (UMA) gate. The circuit showcases a linear depth that favors its scalibility, and we consider the algorithm's ladder of controlled gates a useful pattern for characterizing the inter-core communication workload.

\subsubsection{Grover's main routine}
Grover's algorithm for solving unstructured database search problems is among the first developed quantum algorithms that demonstrates a quadratic speedup compared to its classical counterparts, which makes it a useful subroutine for multiple applications, for instance, power and energy applications (\cite{b59}) and quantum machine learning (\cite{b60}). In essence, the algorithm applies amplitude amplification to increase the probability of measuring the correct answer at the end of the circuit. The main steps of the algorithm are state preparation, the oracle implementation to mark the correct answer depending on the problem input, and the diffusion operator that magnifies the probability of the correct answer before measurement. In this work, we implement the Grover's main routine that starts by initializing the qubit to superposition states and applies iterative instructions to mark the correct answer and amplifies its state amplitude, also referred to as the oracle-diffuser cycle.

\subsubsection{GHZ state}
Entanglement is a key property to unlock the computational power of quantum processors. It is then important to evaluate the modular architecture's support for large entangled states considering the long-range interactions and inter-core communication requirements.
The Greenberger-Horne-Zeilinger state, known as the GHZ state \cite{b61}, is a circuit-size maximally entangled state generated by applying the Hadamard gate on one qubit followed by a ladder of CNOT gates controlled by the qubit in superposition and targeting each of the remaining ones, producing the state
\begin{displaymath}
    \frac{\ket{0}^{\otimes n} +\ket{1}^{\otimes n}}{\sqrt{2}}   
\end{displaymath}
for $n$ qubits. We opted for the circuit implementation where the first qubit is set as the controller since the dependent ladder of CNOT gates serves our interest in monitoring long-range inter-core communications and qubit operations across cores.

\subsubsection{Quantum Fourier Transform}
 The Quantum Fourier Transform (QFT) (\cite{b31}) is a quantum computing routine that applies a discrete Fourier transform to the amplitudes of a wavefunction. It is a fundamental step in several quantum algorithms, including Shor's factoring algorithm (\cite{b62}) and the quantum phase estimation (\cite{b31}). It is implemented utilizing the Hadamard gate, phase shift gates, and controlled phase shift gates. The QFT is proven to be useful for several quantum applications that require large-scale architectures as well, mainly factoring large numbers and solving discrete logarithm problems, which have significant implications in advancing cryptography and computational complexity theory.

\subsubsection{Quantum Approximate Optimization Algorithm}
The Quantum Approximate Optimization Algorithm (QAOA) (\cite{b63}) is a near-term hybrid algorithm designed for solving combinatorial optimization problems. The circuit input state is encoded as a graph where qubits represent the vertices of the graph. By formulating the optimization problem as a mathematical objective function to be minimized or maximized, the algorithm constructs a parameterized quantum circuit where the number of layers $l$ is a tunable parameter and iteratively updates the circuit parameters based on the measured results using classical optimization approaches, aiming to improve the objective function. The best candidate solution is obtained by the end of a fixed number of iterations or until convergence is reached.
For our work, we implement the quantum circuit construction step that performs state preparation and measurement to solve the MaxCut problem of two randomly generated graphs: the Watts Strogatz graph which exhibits short, average path lengths and high clustering (\cite{b64}), and the random Erdos Renyi graph (\cite{b65}) of edge probability $0.2$. The number of vertices corresponds to the total number of qubits. We mainly focus on the circuit mapping and compilation on multi-core quantum devices, rather than solving a given problem. We utilize a circuit ansatz consisting of a sequence of single-qubit rotations and entangling CNOT gates.
    
\subsubsection{Variational Quantum Eigensolver}
The Variational Quantum Eigensolver (VQE) is a hybrid chemistry algorithm designed to determine the ground state energy of a given Hamiltonian (\cite{b66}). The VQE uses a classical optimization algorithm in combination with a parameterized quantum circuit to find the lowest eigenvalue, representing the ground state energy, and its corresponding eigenvector, representing the ground state of a quantum system Hamiltonian, for example a molecule. The algorithm starts with selecting a parameterized circuit that defines a prepared variational state where the number of layers $l$ is a tunable parameter, known as the ansatz. The choice of the ansatz is detrimental to the performance and accuracy of the algorithm. The following step consists of determining the Hamiltonian expectation value by performing measurements on the quantum state obtained from the previous step. Multiple measurements are required to estimate the expectation value accurately. Using a classical optimization algorithm, the parameters of the variational circuit are updated iteratively until the expectation value of the Hamiltonian is minimized. The solution is obtained when a convergence criterion is met, for instance when the expectation value reaches a certain threshold or the parameters converge to a stable value. At this stage, the parameters yielding the minimum expectation value are used to prepare the quantum state that represents the ground state of the Hamiltonian. For our work, we implement two different types of Hardware-efficient ansatze, the first displaying a sequential ladder of CNOT gates as depicted in (\cite{b66}), and the second showcasing parallel execution of CNOT gates as shown in (\cite{b67}).

\subsection{Performance metrics for multi-core quantum processors} \label{metrics}
Considering classical distributed systems, multiple performance metrics have been used to assess and optimize of NoCs for classical modular computers. Studying the temporal distribution of traffic patterns, researchers and engineers can develop strategies to predict and mitigate congestion before it impacts system performance. Spatial hotspots, or areas with consistently high traffic, may indicate potential failure points in the network due to overheating or overuse. By identifying hotspots, network designers may optimize the network topology and routing algorithms to distribute traffic more evenly, preventing any single node or link from becoming a bottleneck.
 
In analogy to spatio-temporal characterization of the traffic in distributed classical systems, we evaluate the computational capacity of quantum multi-core processors for executing our selected set of circuits according to several metrics that represent the computational performance and communication requirements. The introduced metrics characterize the communication overhead in modular architectures, and potentially the compilation process and processor technology. 
    
\subsubsection{Computation-to-Communication Ratio (CCR)}
Executing quantum circuits on modular architectures demands implementing additional inter-core operations according to a given communication protocol, which presumably delays the computational tasks. Designing robust processors that achieve an efficient computation process requires a trade-off for allocating computation and communication qubits. The inter-core communication protocol we employ for mapping circuits on modular architectures is the teleportation protocol. We define the computation-to-communication ratio (CCR) as the normalized ratio of average single- and two-qubit operations $n_{ops}$ to the average number of teleportation operations $n_{telep}$ per qubit:
\begin{equation}
\mathbf{CCR} = \dfrac{n_{ops}-n_{telep}}{n_{ops}+n_{telep}}
\end{equation}
If the ratio tends to 1, we conclude that computation tasks are dominating the execution process. If it tends to -1 then communications are dominating.
  
\subsubsection{Resource usage uniformity in space and time}
\vspace{0.1cm}
\noindent \textbf{Mean qubit hotspotness:} We investigate the qubit hotspotness, which is related to the number of gates applied on physical qubits, to estimate the distribution uniformity of operations per qubit throughout the system. We account for both computation and teleportation operations. We define the mean qubit hotspotness as the ratio of the variance of operations $\sigma^2_{ops/qubit}$ to the average operations applied to a physical qubit $\mu_{ops/qubit}$ following the compilation process: 
    \begin{equation}
    \overline{H}_{qubit} = \dfrac{\sigma^2_{ops/qubit}}{\mu_{ops/qubit}}
    \end{equation}  

\vspace{0.1cm}
\noindent \textbf{Mean core hotspotness:} For an optimal usage of resources of the processor, the computation and communication operations need to be uniformly distributed across the cores as per the circuit requirements. We look into the core hotspotness during the program execution to identify the cores that operate inter-core communications and are mostly involved in the communication operations. We correlate the core hotspotness to the distribution of inter-core communications in the system and define the mean core hotspotness as the ratio of the variance of teleportation operations $\sigma^2_{telep/core}$ to the average teleportation operations $\mu_{telep/core}$ for each core:
    \begin{equation}
    \overline{H}_{core} = \dfrac{\sigma^2_{telep/core}}{\mu_{telep/core}}
    \end{equation}
    
\vspace{0.1cm}
\noindent \textbf{Longest gate sequence:} Since qubits are highly susceptible to state disruptions due to gate application, performing a long sequence of gates would presumably result in deteriorating the computational process. As we scale the circuit size, we look into the increment of operations executed. For our analysis, we determine the longest gate sequence as the maximum number of gates applied to a qubit in a given circuit:
    \begin{equation}
    G = \max_{\forall q} |G_{ij}|
    \end{equation}
where $G_{ij}$ is the set of gates operating qubits $i$ and $j$.

Estimating the number of operations needed to execute a given circuit provides insight into the capacity of quantum devices to support large-scale computations within the limits of the qubit technology. 
    
\vspace{0.1cm}
\noindent \textbf{Qubit lifespan:} This characterizes the capacity of the system to execute large-scale circuits within the limits of qubit technology and decoherence times. We interpret the qubit lifespan as the longest timeslice length that separates the first and the last gate applied to a virtual qubit in a given circuit over the execution timeslices:
    \begin{equation}
    L = \max_q {L_q}
    \end{equation}
where
    \begin{equation*}
    L_q = \max_{q \in \{i,j\}} T_{exec}(G_{ij}) - \min_{q \in \{i,j\}} T_{exec}(G_{ij}) 
    \end{equation*}
such that $T_{exec}(g)$ corresponds to the execution timeslice of a specific gate $g$. 
\footnote{We define a timeslice as a discrete segment of time during which a set of quantum gates are applied, and its duration is determined by the longest gate applied.}

\vspace{0.1cm}
\noindent \textbf{Temporal burstiness:} We look into the inter-core communication burstiness throughout the program execution for the purpose of monitoring the teleportation clustering in timeslices. The system is supposed bursty if there exists timeslices exhibiting a high number of teleportation operations clustered subsequently, followed by timeslices of low or spaced inter-core communications. This implies a non-uniform distribution of communications over runtime and directly impacts the performance of the system, resource allocation, and operation scheduling. We characterize the burstiness of a quantum system as the ratio of the variance of parallel teleportations $\sigma^2_{telep(t)}$ to the mean of parallel teleportations throughout the execution timeslices $\mu_{telep(t)}$:
    \begin{equation}
    \overline{B} = \dfrac{\sigma^2_{telep(t)}}{\mu_{telep(t)}}
    \end{equation}         
        
\subsubsection{Locality}

It is crucial to characterize the qubits' spatial and temporal locality for the purpose of evaluating the circuit partitioning techniques employed at the compiler level.
    
\vspace{0.1cm}
\noindent \textbf{Qubit temporal locality:} We quantify the qubit temporal locality as the total number of teleportation required for a given circuit execution:
    \begin{equation}
        T = \sum_{ts} T_{ts}
    \end{equation}
where $T_{ts}$ is the number of teleportation per timeslice. 

\vspace{0.1cm}
\noindent \textbf{Qubit spatial locality:} The qubit spatial locality refers to the average duration that a logical qubit stays in a specific core, indicating the frequency of inter-core qubit movements. A higher average stay-time implies qubits remain relatively stationary, thus minimizing disturbance of their states. However, a lower average stay-time indicates frequent movement of qubits, which increases the risk of state loss.
We define the qubit spatial locality as the average qubit stay-time in a core over the program execution time:
    \begin{equation}
    S = \frac{d_{qubit/core}}{t_{exec}}
    \end{equation}
 where $d_{qubit/core}$ is the average stay-time of a qubit in a core and the $t_{exec}$ is the total execution time defined in timeslices.

\section{Results} \label{results}

\subsection{Impact of circuit mapping}

\begin{figure*}[h]
\begin{subfigure}[t!]{0.5\textwidth}
 \centering
\captionsetup{justification=raggedright,singlelinecheck=false}
\raisebox{1.2cm}{\includegraphics[scale=0.35]{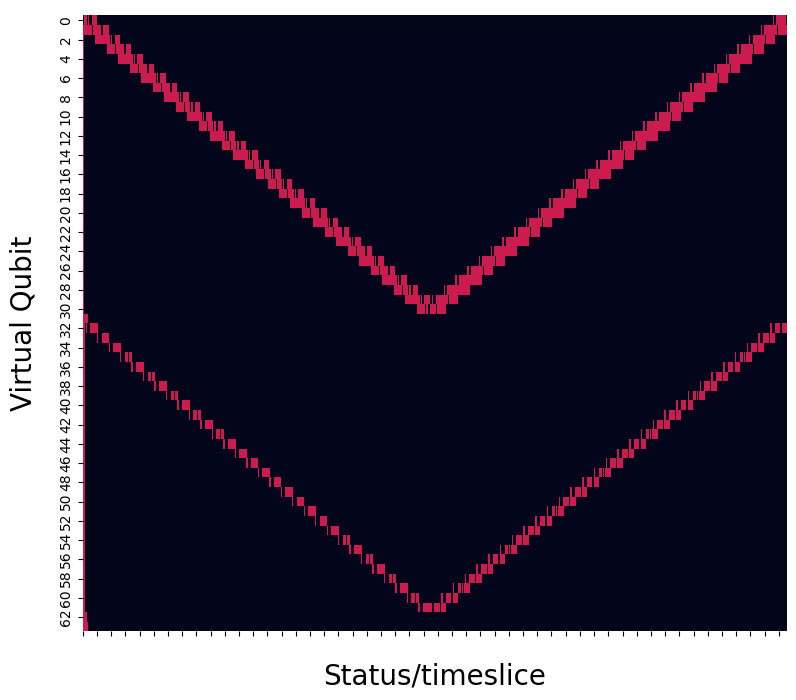}}
 \vspace{0.2cm}
\caption{The Cuccaro Adder circuit logical structure for 64 qubits. Computation times are represented in red.}
 \label{virt_cuccaro}
\end{subfigure}%
\begin{subfigure}[t!]{0.5\textwidth}
 \centering
\captionsetup{justification=raggedright,singlelinecheck=false}
 \includegraphics[scale=0.35]{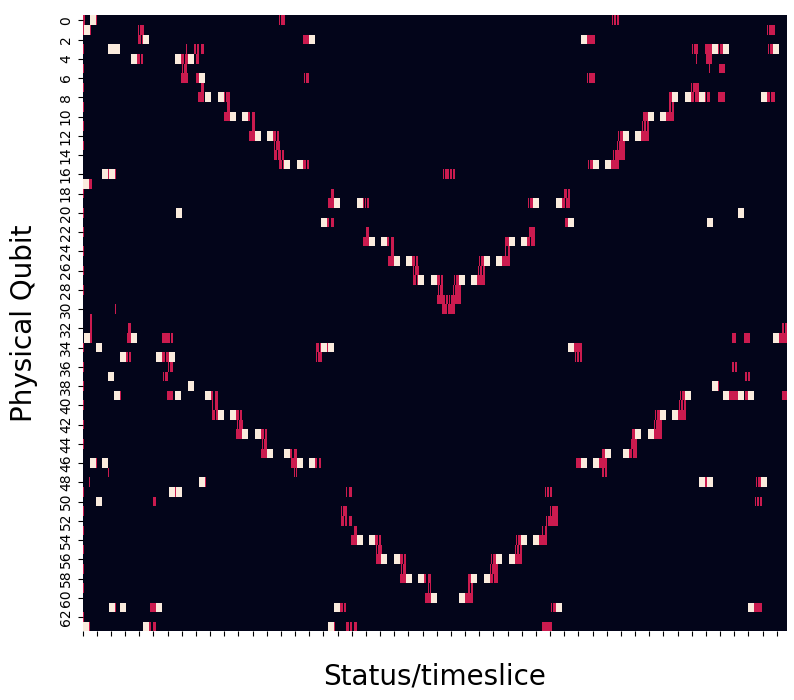}
 \vspace{0.3cm}
 \caption{The Cuccaro Adder circuit mapping to physical qubits on a modular architecture of 16 qubits $\times$ 4 cores with all-to-all connectivity. Computation, communication, and idling times are represented in red, white, and black, respectively.}
 \label{phys_cuccaro}
\end{subfigure}%
\caption{The Cuccaro Adder circuit mapping to virtual and physical qubits.}
  \label{cuccaro_circuits}
\end{figure*}

In Figure \ref{virt_cuccaro}, we show the logical structure of the Cuccaro Adder input circuit representing the interacting qubits and operation times throughout the execution timeslices. On the other hand, the physical mapping trace in Figure \ref{phys_cuccaro} showcases the Cuccaro Adder circuit mapping to physical qubits on a modular architecture and displays computation, communication and idling qubits and their respective operational times. This presents a general overview and visualization of the necessary resources to be allocated for a program execution considering a given architecture.

Circuit mapping onto the physical architecture as performed by the compiler imposes several changes to the circuit logical structure. These structural changes exhibit the application of the mapping algorithm which aims to move qubits across processors in order to place interacting qubits into adjacent positions. Correspondingly, allocating more qubits into a single core would reduce the amount of teleportations required as local qubit interactions are favoured. While this setup would mitigate the communication overhead, it results in imposing the architectural constraints originating from a densely-packed quantum processor, as introduced beforehand. We present the circuit mapping traces of the various executed circuits in Appendix \ref{traces}. Additionally, we state that the circuit structure and qubit interactions are essentially algorithm dependent: the QAOA\_02 and QAOA\_WS circuits, for instance, implement the same ansatz yet different input graphs. Consequently, the corresponding virtual traces are entirely different, indicating distinct circuit structures. This also impacts the number of gates applied and the inter-core operations required.

Circuits characterized with short-range intra- and inter-core gates where two-qubit operations are applied on nearest-neighbor or localized qubits, i.e qubits placed on the same core, such as Cuccaro, Grover, and QAOA\_WS, and VQE\_HEA\_1 tend to preserve their structures after the mapping process, indicating that a relatively lower number of qubit movements were required to comply with an optimal mapping to physical qubits. The given architecture and the circuit partitioning algorithm are therefore advantageous for the execution of these circuits.

For algorithms that require long-range or extensive parallel execution of two-qubit gates, such as for GHZ, QFT, QAOA\_02 and VQE\_HEA\_2, the compiler imposes several rearrangements of qubit placement to comply with the input instructions. We note scattered communication and computation operations in their corresponding physical circuits. The GHZ circuit is a particular example of sequential two-qubit gates applied from one qubit to all others, and therefore exhibits both localized, short-range and cross-core, long-range operations, which require extensive inter-core qubit movement. In the physical mapping, we show that the compiler moves the target qubits to the first core where the qubit 0 is located in order to enable direct application of CNOTs. This appoints a frequent qubit movement and serial application of teleportation instructions. 

Generally, a relatively limited number of qubits per chip contributes to the surging resource requirement as we increase the number of cores, which calls for a trade-off of the number of qubits per chip and the total circuit size. The mapping process gets more complex with a large number of cores, particularly for circuits exhibiting a high number of gates such as the QFT and QAOA\_02 instances. This presumably gives an explanation for the scattered operation patterns and noticeable transformations from virtual to physical mapping.

\subsection{Strong scaling}

\begin{figure*}[htbp] 

    \begin{subfigure}[t!]{0.25\textwidth}
     \centering
     \includegraphics[scale=0.31]{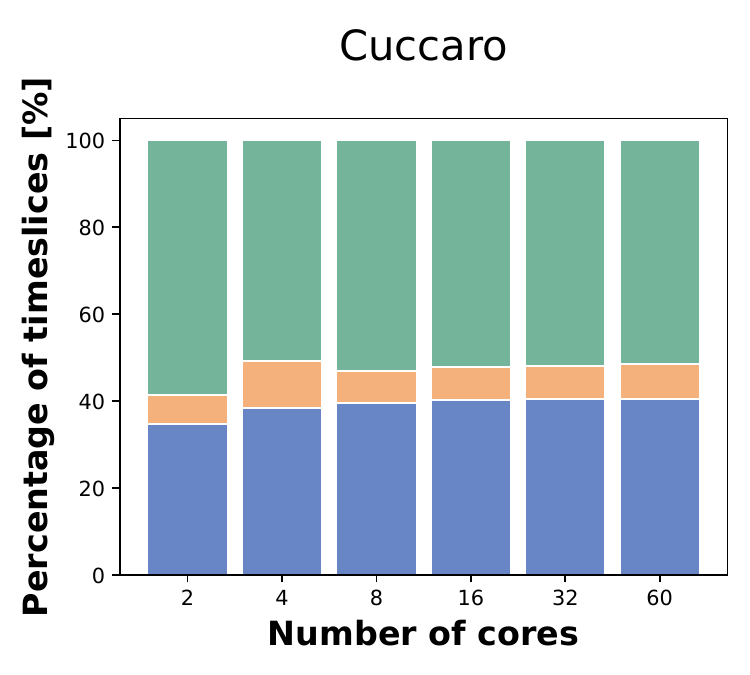}
    \end{subfigure}%
    \begin{subfigure}[t!]{0.25\textwidth}
     \centering
     \includegraphics[scale=0.31]{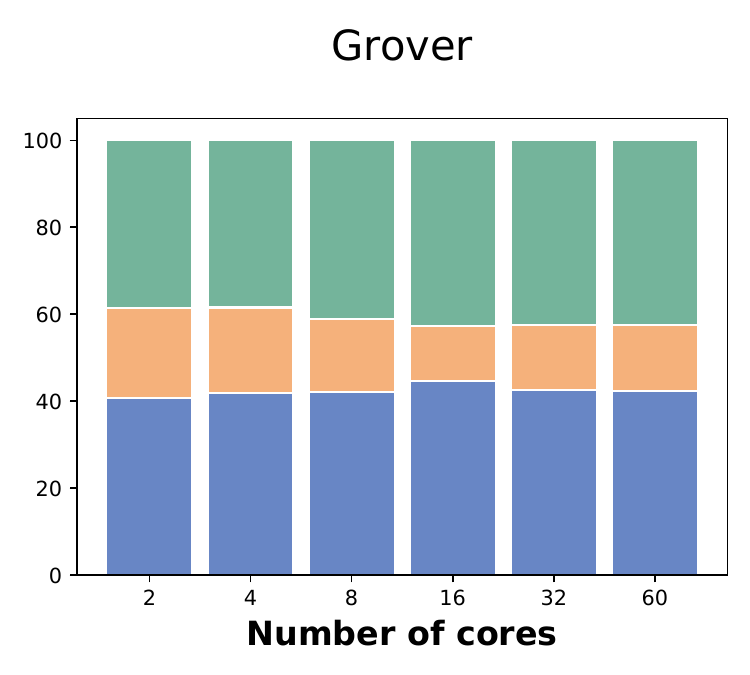}
    \end{subfigure}%
    \begin{subfigure}[t!]{0.25\textwidth}
     \centering
     \includegraphics[scale=0.31]{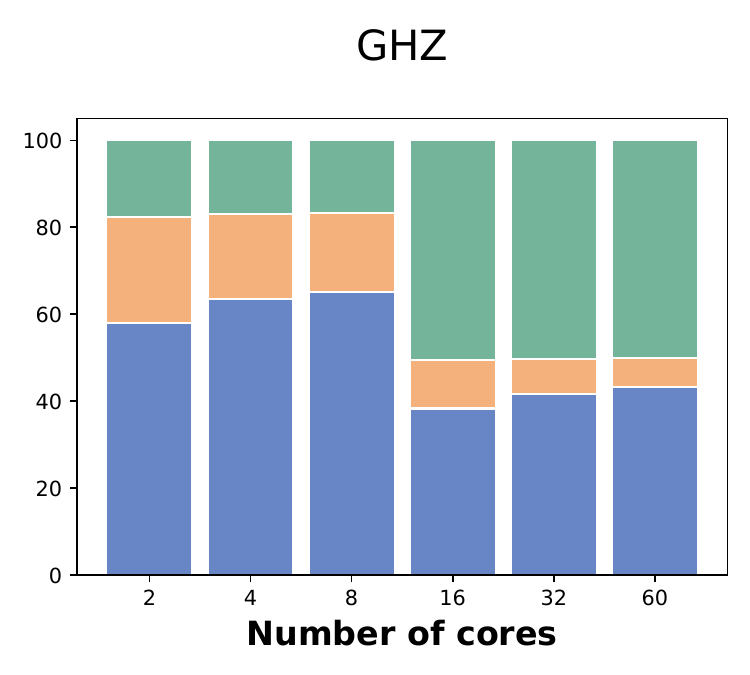}
    \end{subfigure}%
    \begin{subfigure}[t!]{0.25\textwidth}
     \centering
     \includegraphics[scale=0.31]{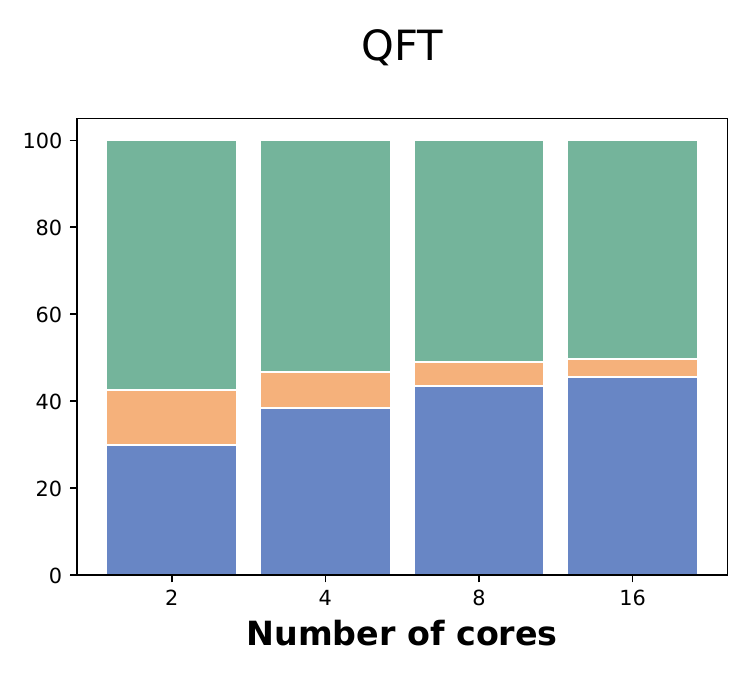}
    \end{subfigure} 
    \begin{subfigure}[t!]{0.25\textwidth}
     \centering
     \includegraphics[scale=0.31]{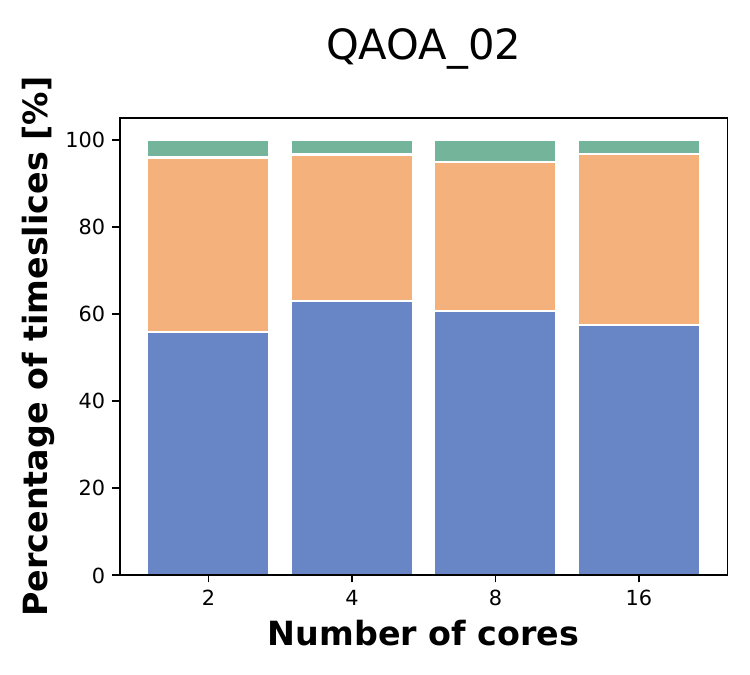}
    \end{subfigure}%
    \begin{subfigure}[t!]{0.25\textwidth}
     \centering
     \includegraphics[scale=0.31]{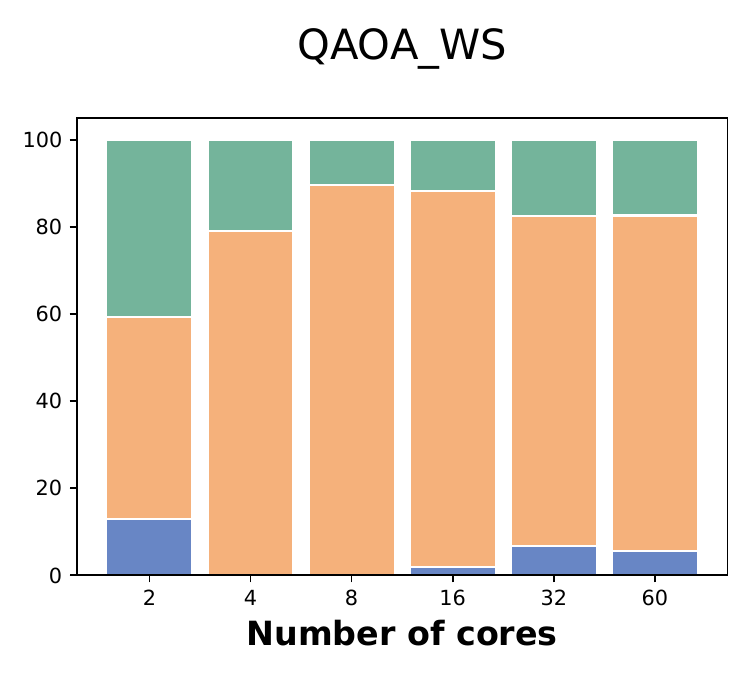}
    \end{subfigure}%
    \begin{subfigure}[t!]{0.25\textwidth}
     \centering
     \includegraphics[scale=0.31]{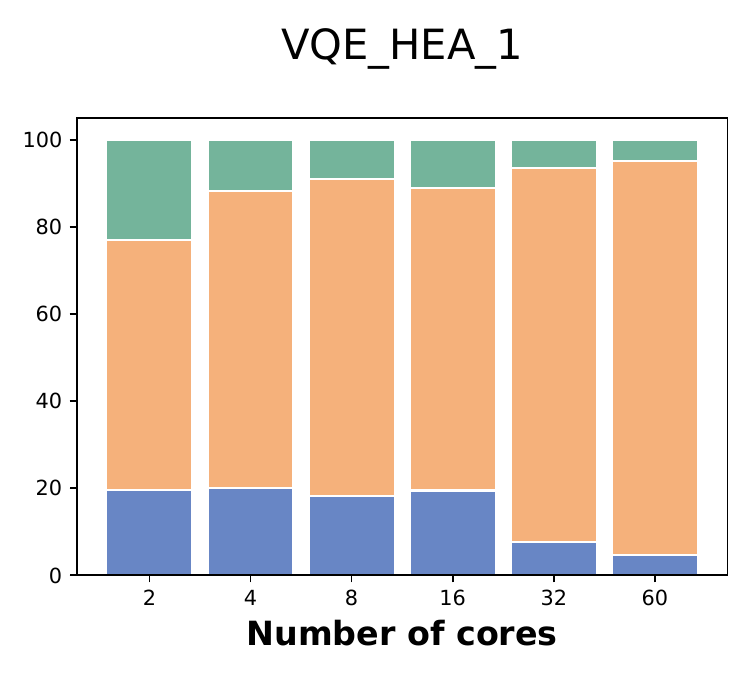}
    \end{subfigure}%
    \begin{subfigure}[t!]{0.25\textwidth}
     \centering
     \includegraphics[scale=0.31]{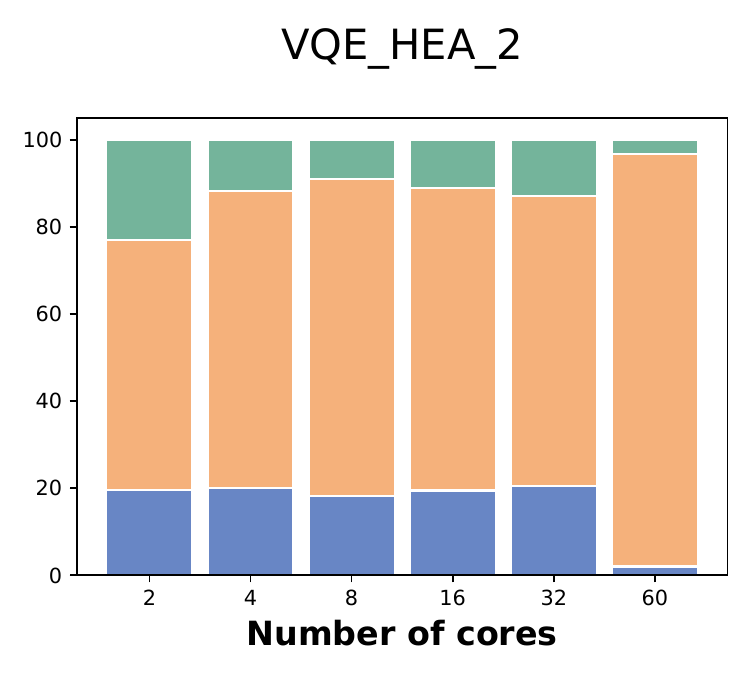}
    \end{subfigure}%
    \vspace{0.3cm}
    
    \caption{Distribution of computation and communication operations over the execution timeslices in the strong scaling. Blue, orange, and green indicate the communication, parallel communication and computation, and computation operations, respectively.}
      \label{ccr_strong_scaling}
      
\end{figure*}

\begin{figure*}[htbp]  
\centering
\begin{subfigure}[t]{0.5\textwidth}
  \centering
  \includegraphics[scale=0.32]{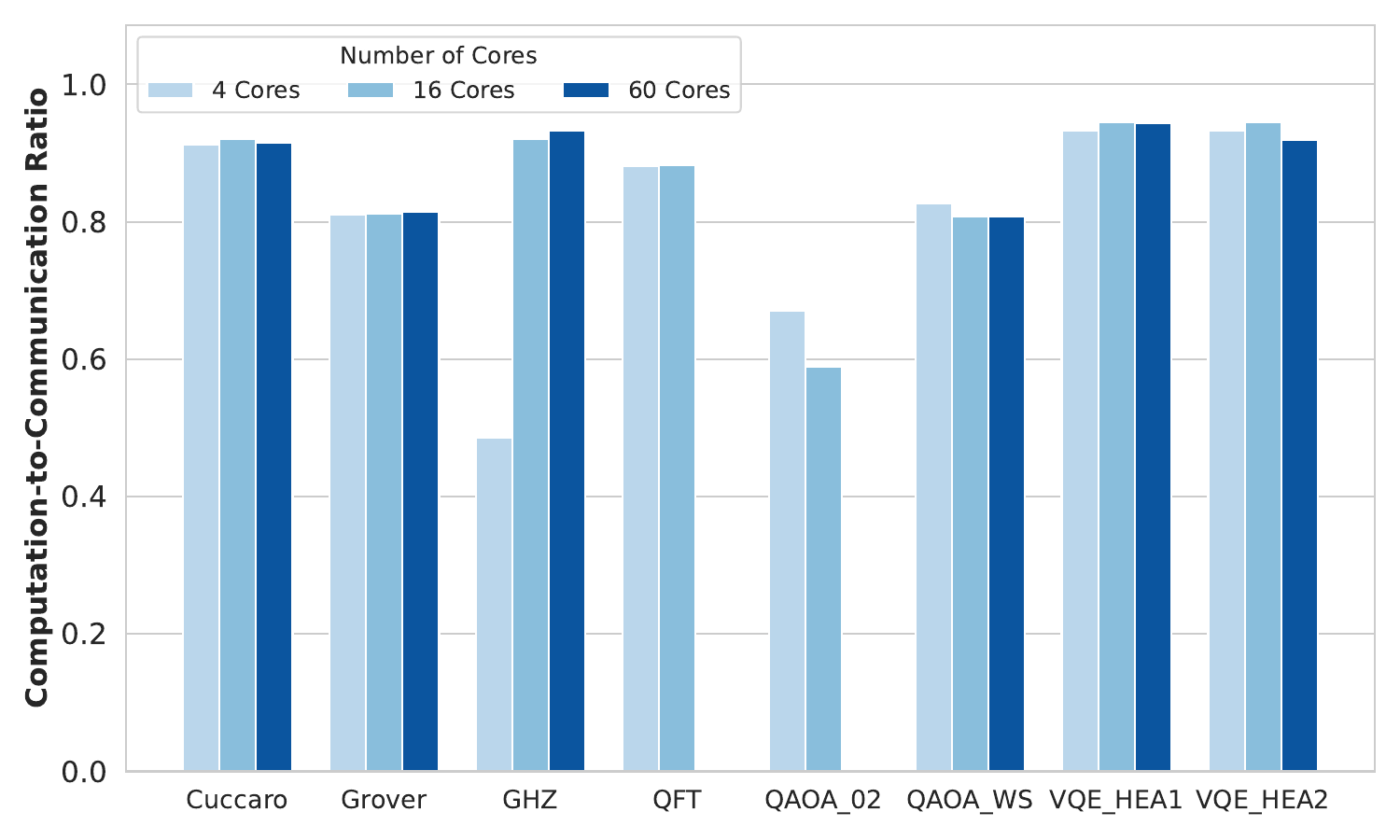}
\end{subfigure}%
\begin{subfigure}[t]{0.5\textwidth}
  \centering
  \includegraphics[scale=0.32]{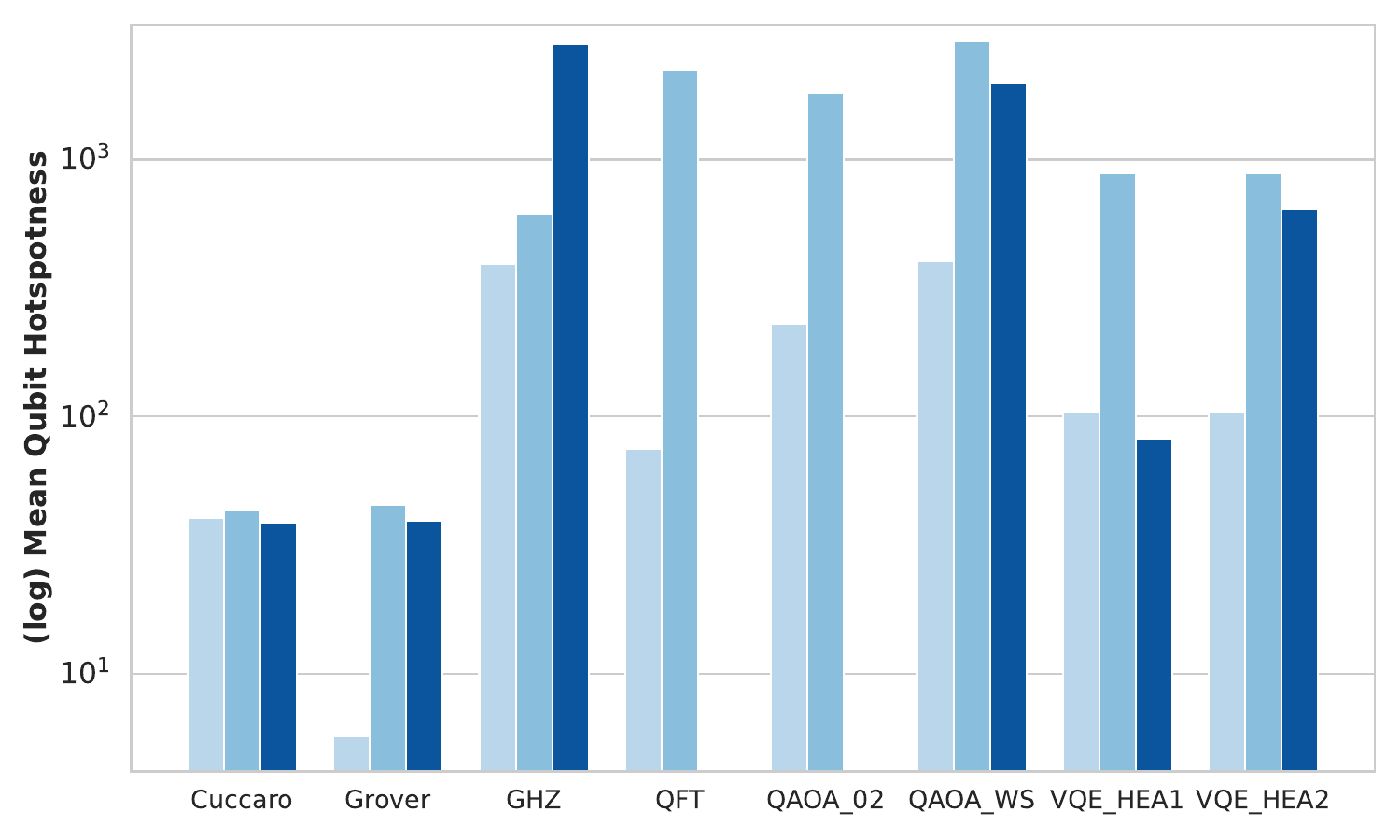}
\end{subfigure}%
\vspace{0.3cm}
\begin{subfigure}[t]{0.5\textwidth}
  \centering
  \includegraphics[scale=0.32]{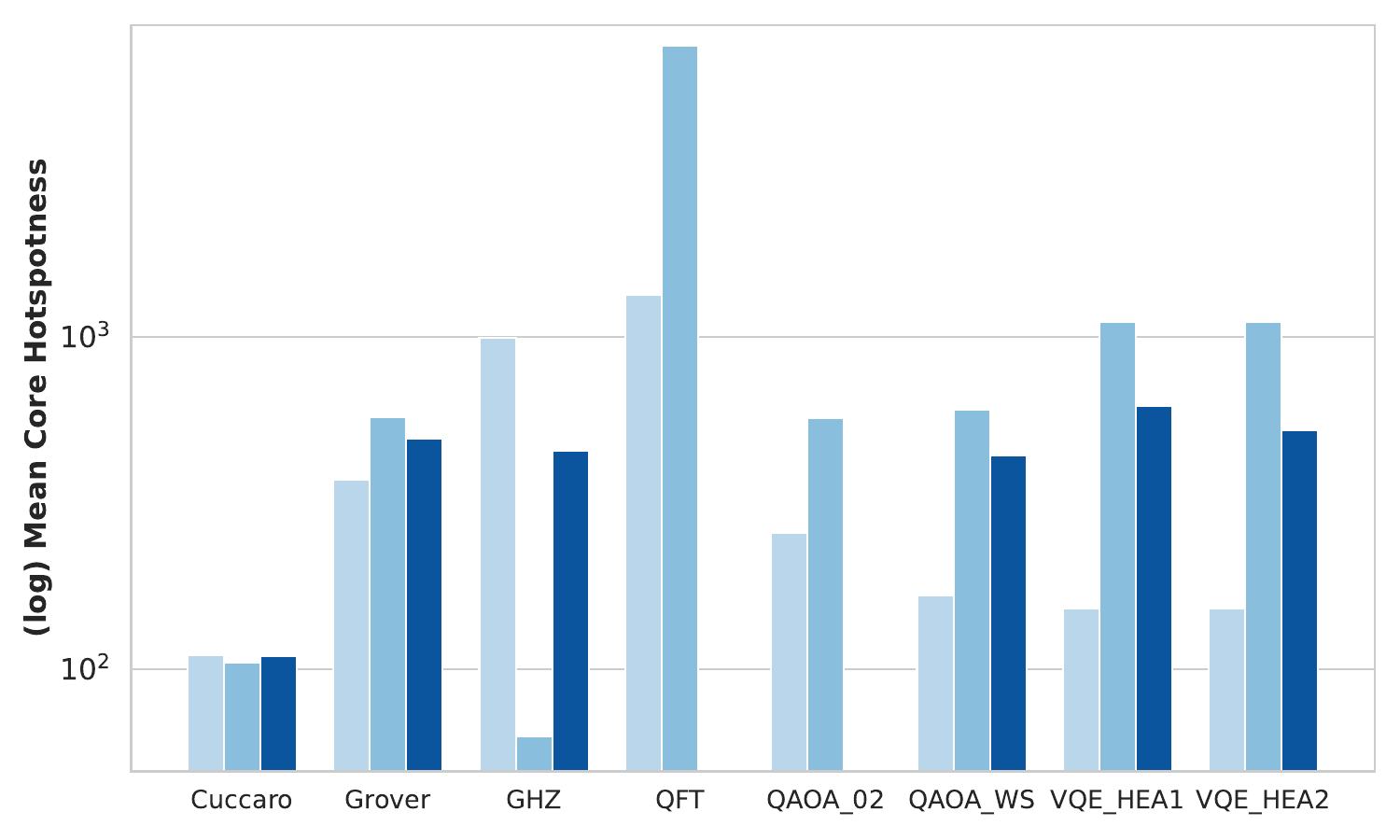}
\end{subfigure}%
\begin{subfigure}[t]{0.5\textwidth}
  \centering
  \includegraphics[scale=0.32]{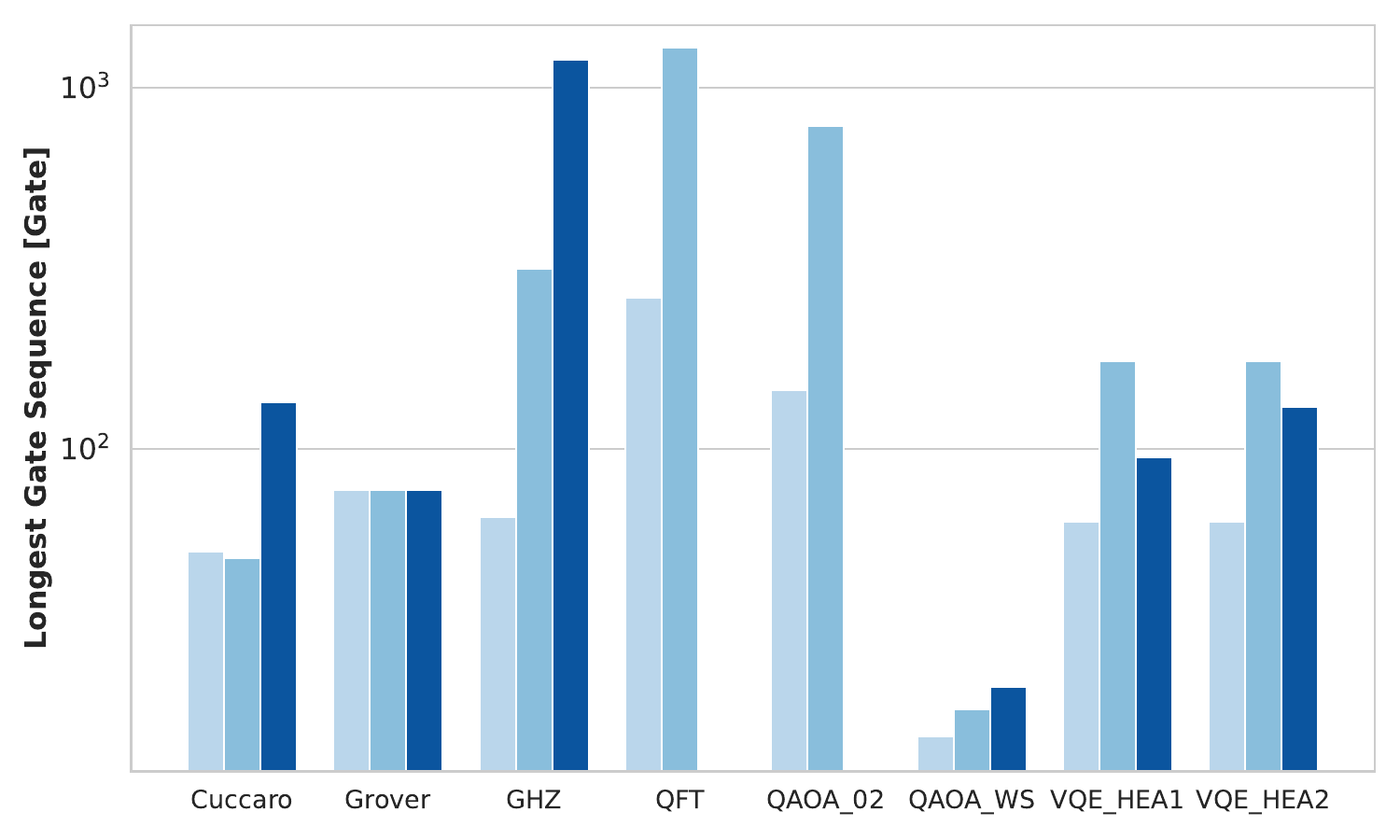}
\end{subfigure}%
\vspace{0.3cm}
\begin{subfigure}[t]{0.5\textwidth}
  \centering
  \includegraphics[scale=0.32]{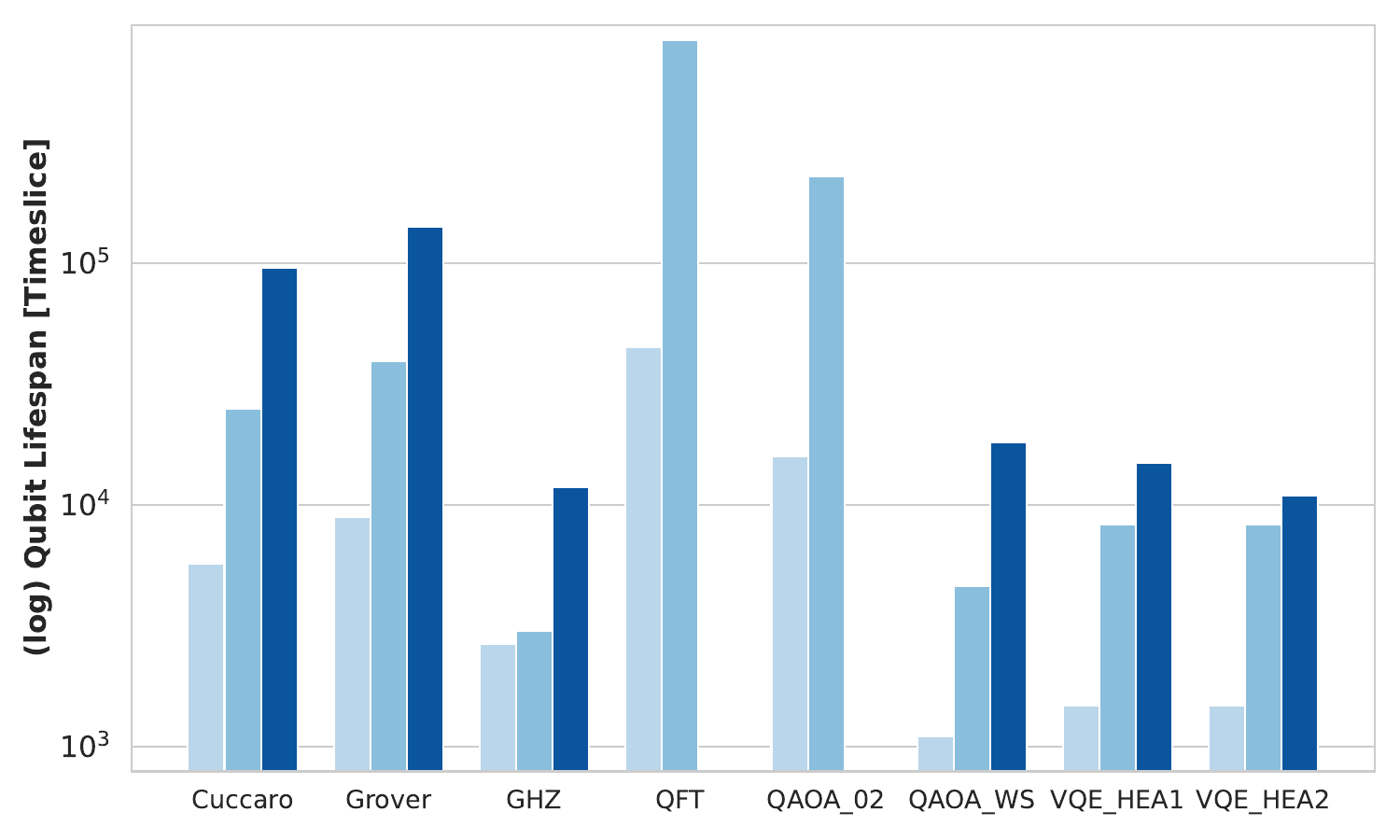}
\end{subfigure}%
\begin{subfigure}[t]{0.5\textwidth}
  \centering
  \includegraphics[scale=0.32]{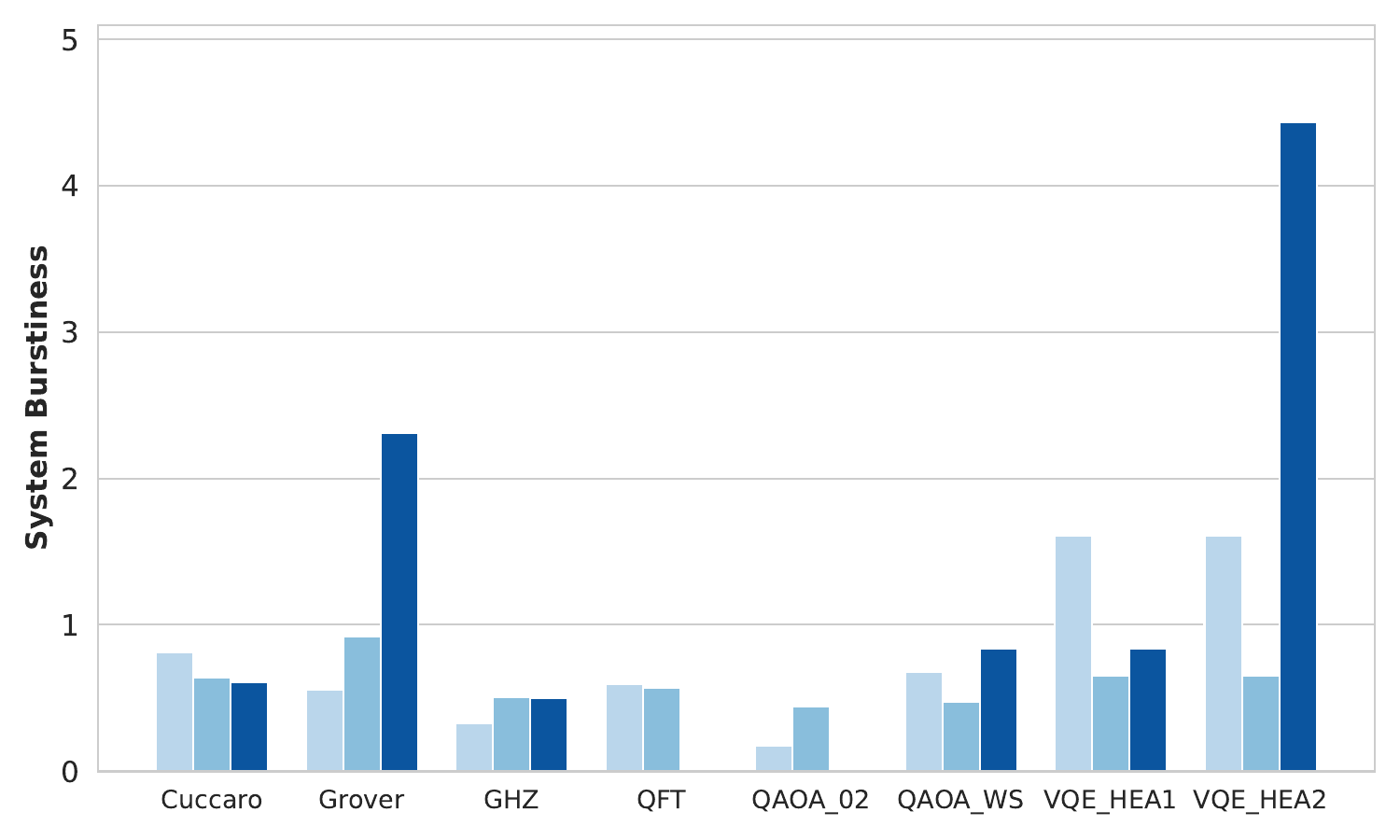}
\end{subfigure}%
\vspace{0.3cm}
\begin{subfigure}[t]{0.5\textwidth}
  \centering
  \includegraphics[scale=0.32]{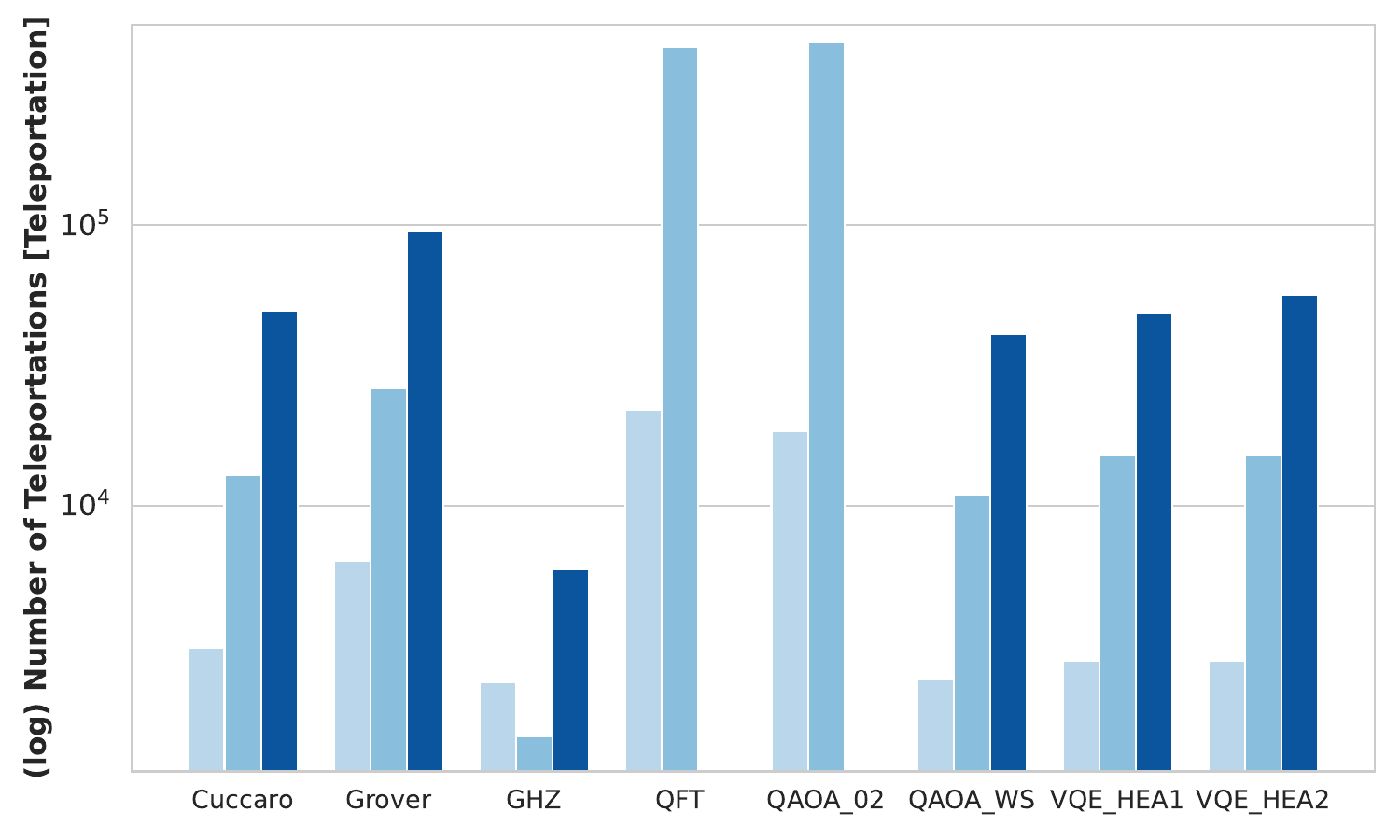}
\end{subfigure}%
\begin{subfigure}[t]{0.5\textwidth}
  \centering
  \includegraphics[scale=0.32]{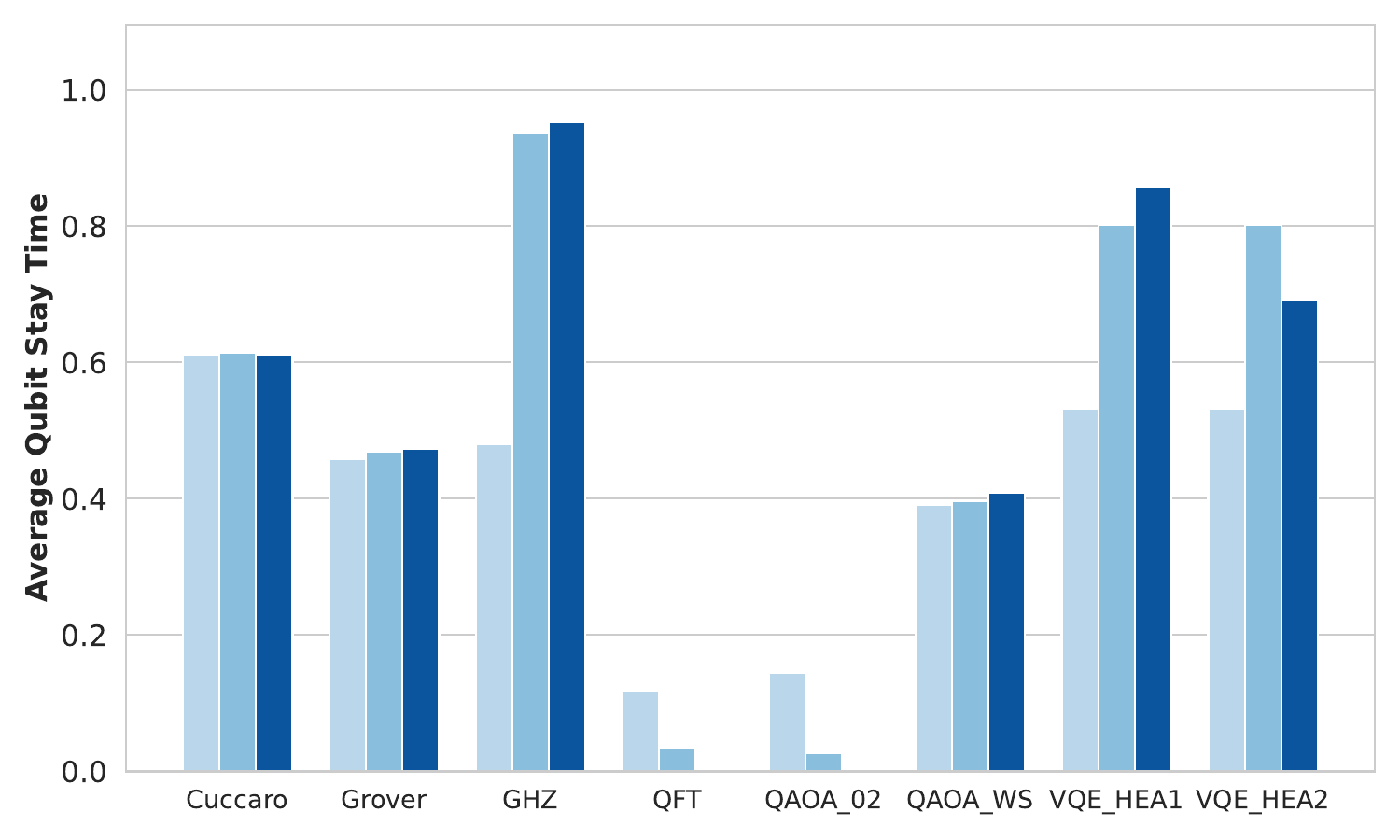}
\end{subfigure}%
\vspace{0.3cm}
\caption{Performance metrics for the selected algorithms executed on modular architectures of 4, 16, and 60 cores, respectively, supporting around 1000 qubits in the strong scaling.}
\label{strong_scaling}
\end{figure*}

\begin{figure*}[htbp] 

    \begin{subfigure}[t!]{0.25\textwidth}
     \centering
     \includegraphics[scale=0.31]{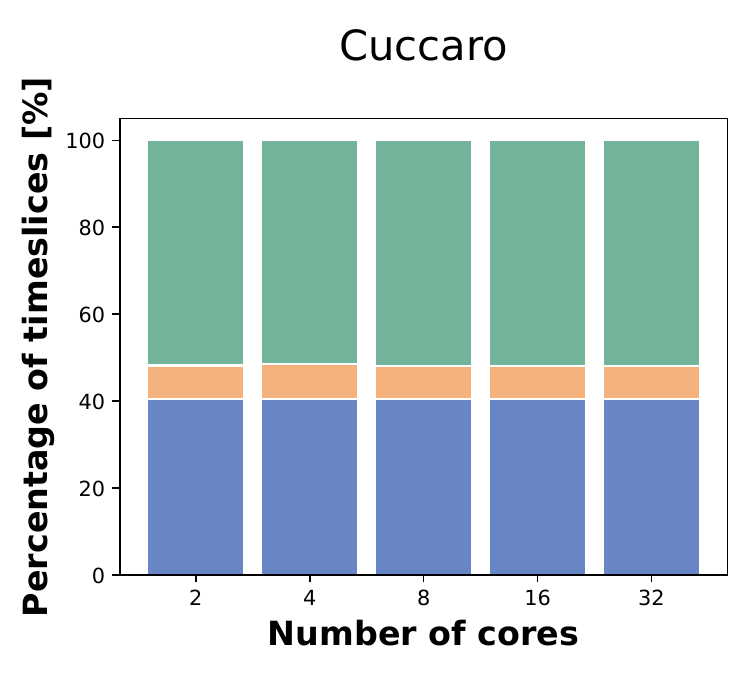}
    \end{subfigure}%
    \begin{subfigure}[t!]{0.25\textwidth}
     \centering
     \includegraphics[scale=0.31]{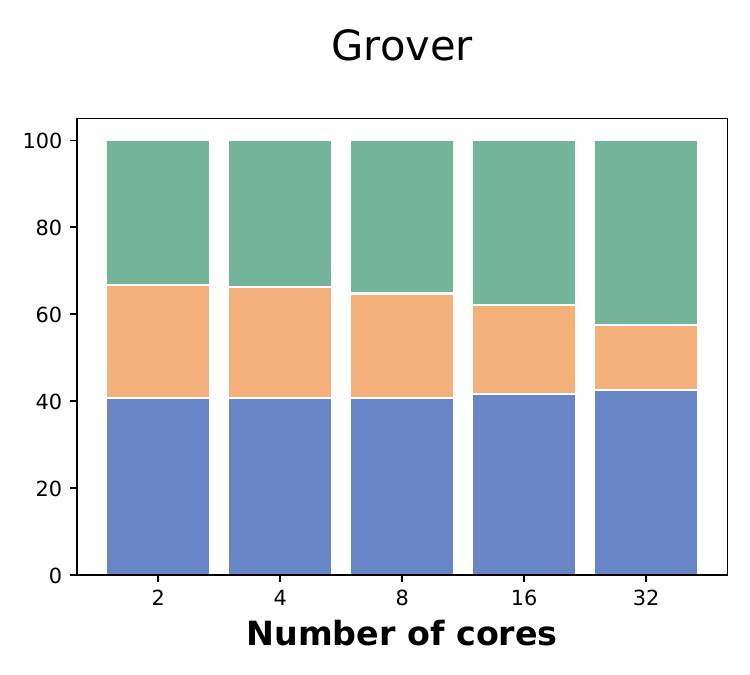}
    \end{subfigure}%
    \begin{subfigure}[t!]{0.25\textwidth}
     \centering
     \includegraphics[scale=0.31]{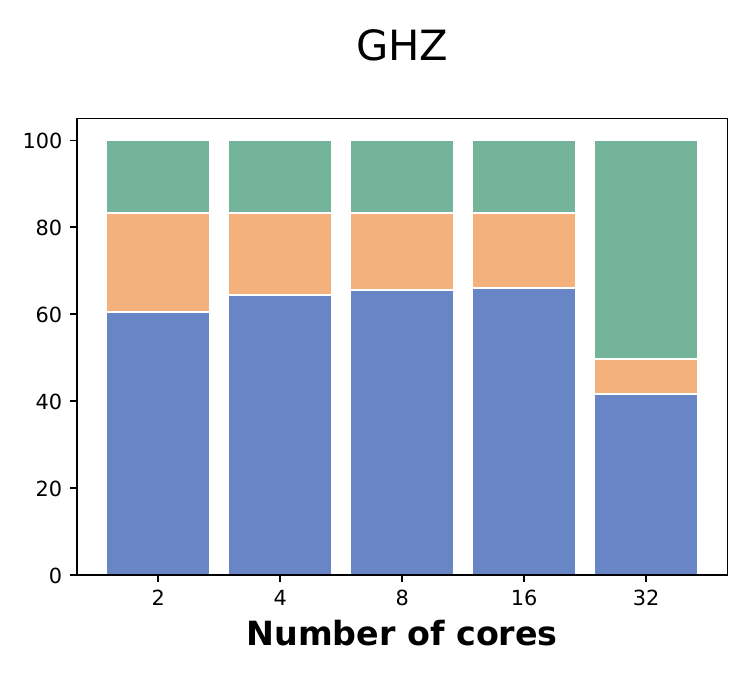}
    \end{subfigure}%
    \begin{subfigure}[t!]{0.25\textwidth}
     \centering
     \includegraphics[scale=0.31]{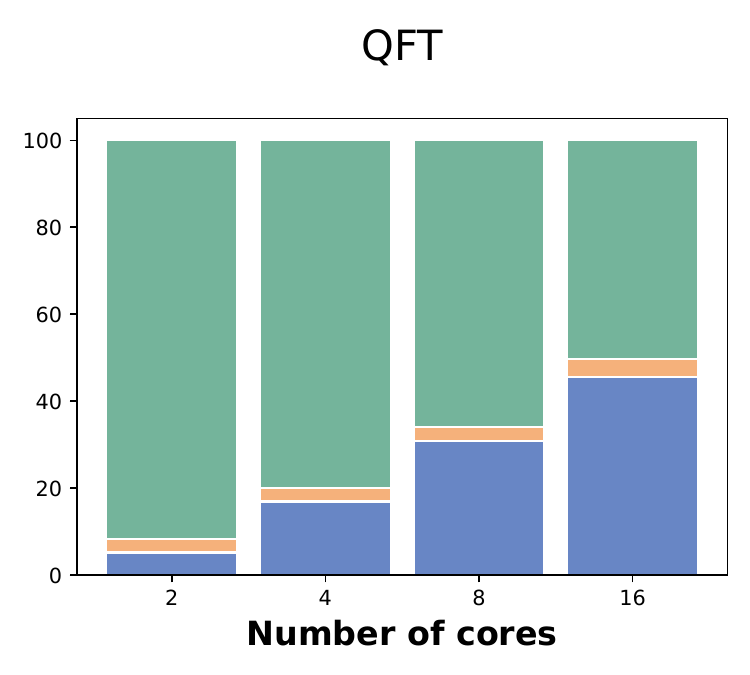}
    \end{subfigure} 
    \begin{subfigure}[t!]{0.25\textwidth}
     \centering
     \includegraphics[scale=0.31]{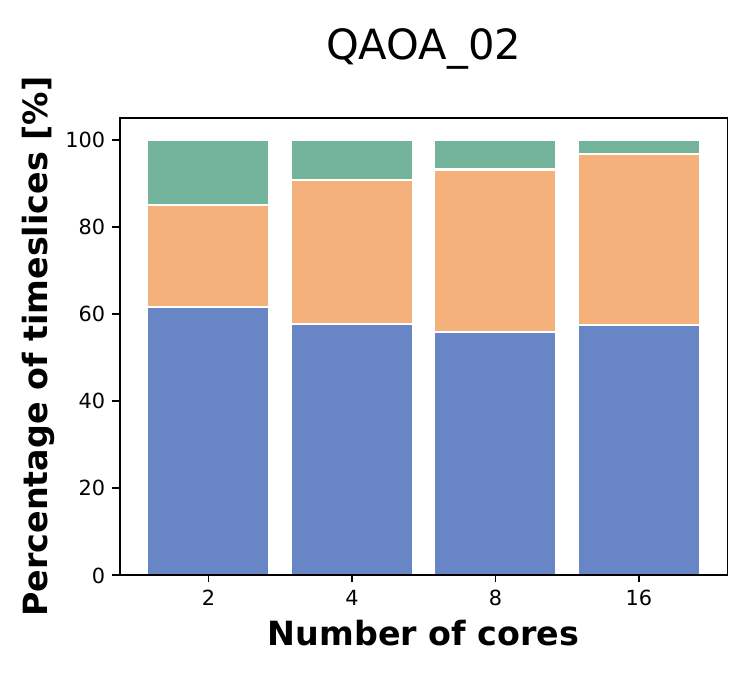}
    \end{subfigure}%
    \begin{subfigure}[t!]{0.25\textwidth}
     \centering
     \includegraphics[scale=0.31]{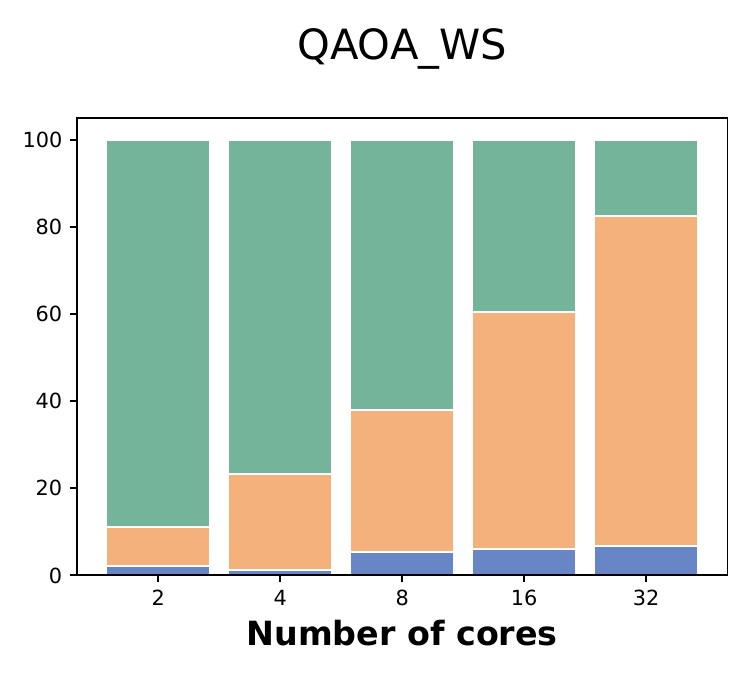}
    \end{subfigure}%
    \begin{subfigure}[t!]{0.25\textwidth}
     \centering
     \includegraphics[scale=0.31]{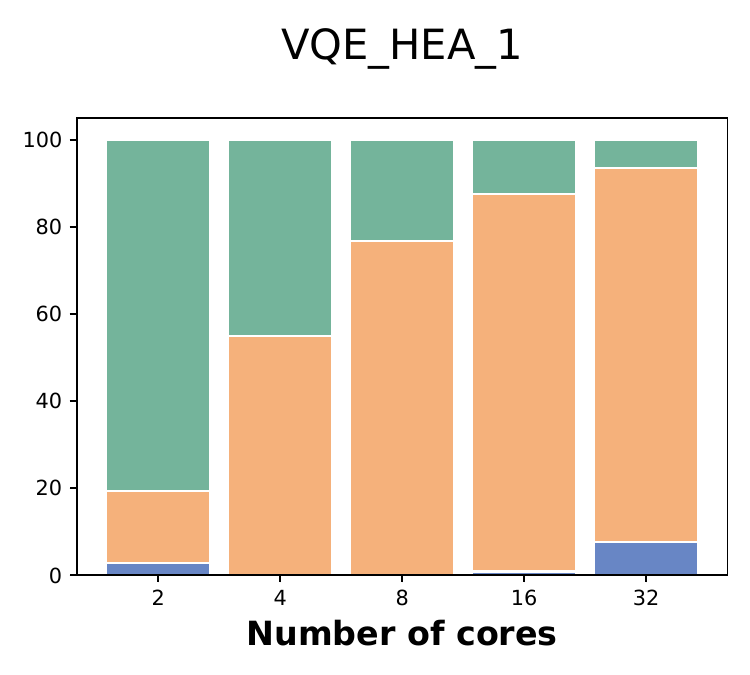}
    \end{subfigure}%
    \begin{subfigure}[t!]{0.25\textwidth}
     \centering
     \includegraphics[scale=0.31]{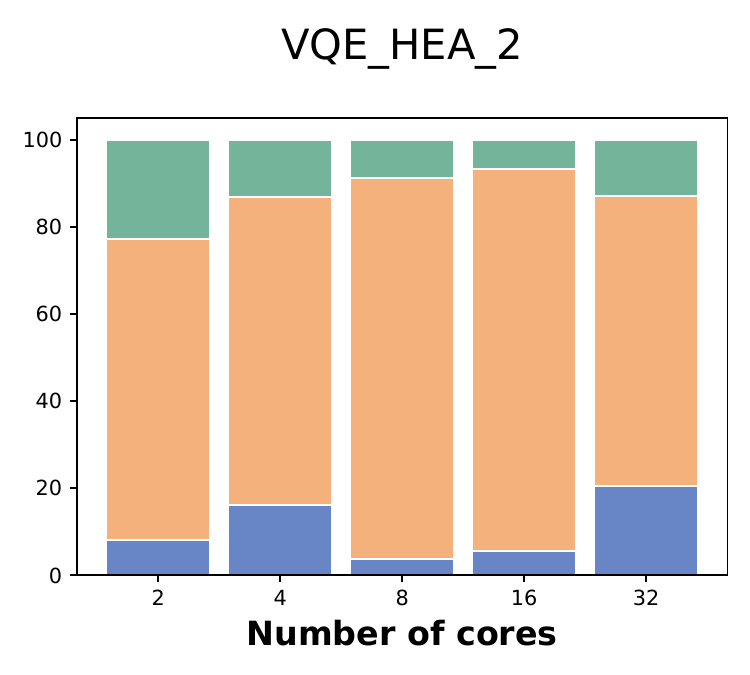}
    \end{subfigure}%
    \vspace{0.3cm}
    
    \caption{Distribution of computation and communication operations over the execution timeslices in the weak scaling. Blue, orange, and green indicate the communication, parallel communication and computation, and computation operations, respectively.}
      \label{ccr_weak_scaling}
      
\end{figure*}

\begin{figure*}[htbp]  
\centering
\begin{subfigure}[t]{0.5\textwidth}
  \centering
  \includegraphics[scale=0.32]{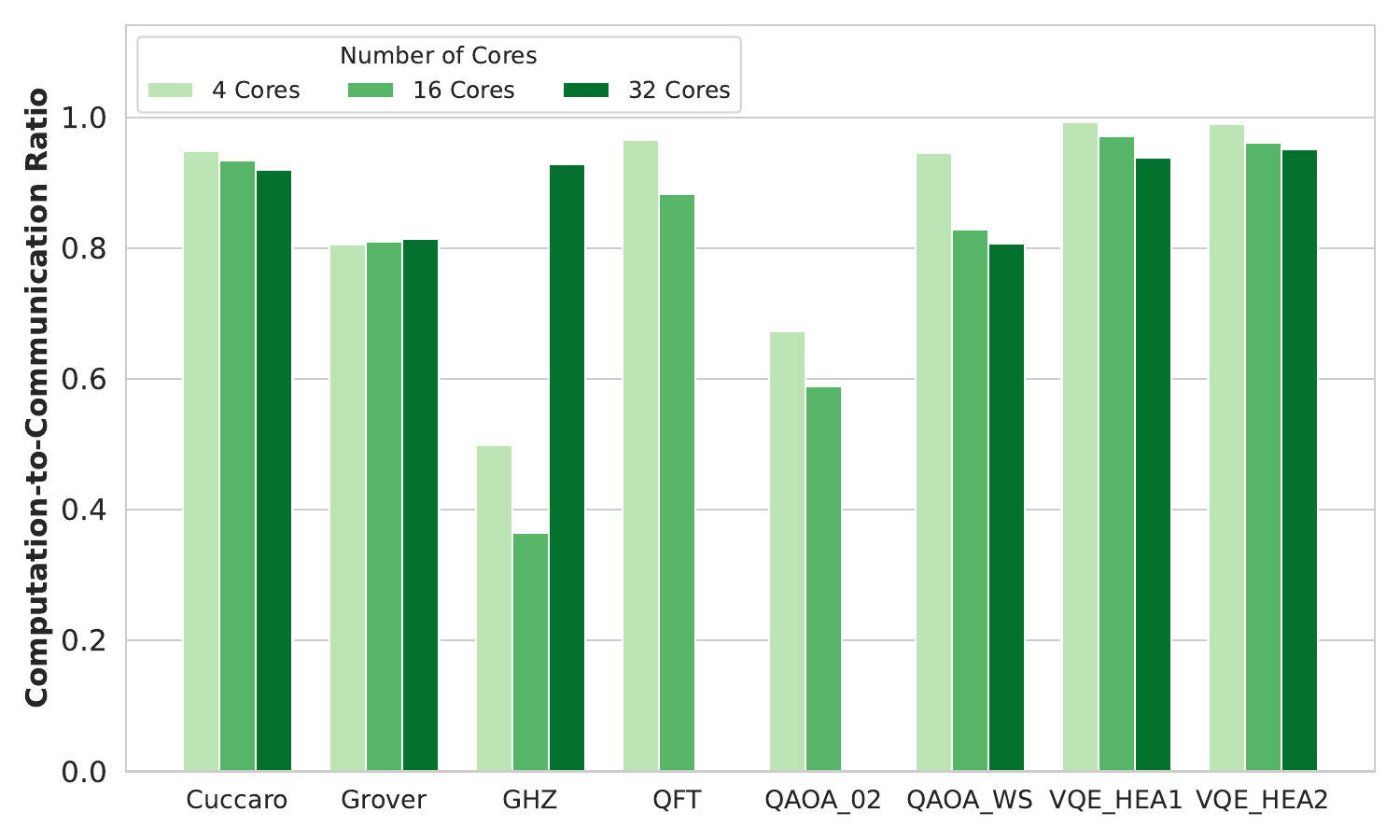}
\end{subfigure}%
\begin{subfigure}[t]{0.5\textwidth}
  \centering
  \includegraphics[scale=0.32]{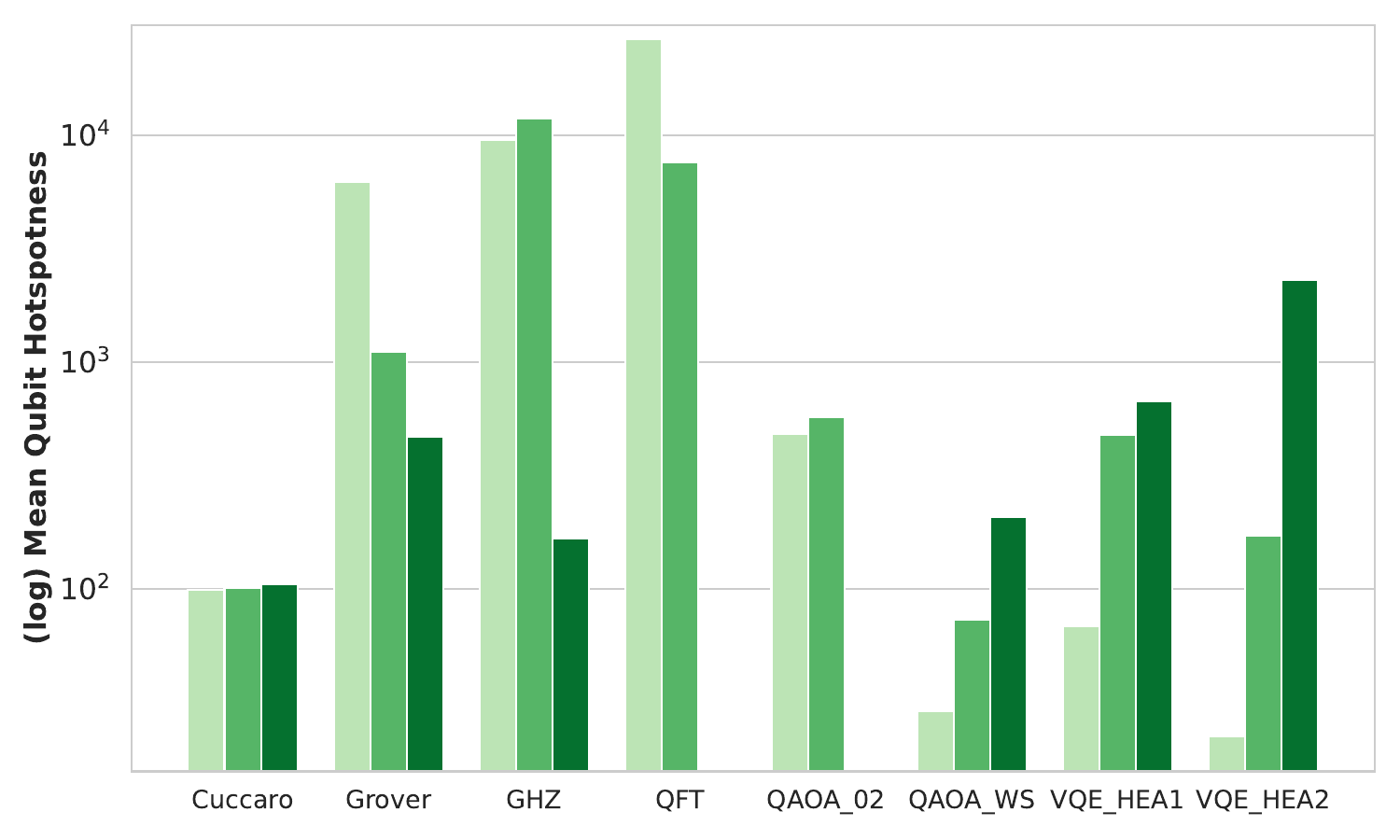}
\end{subfigure}%
\vspace{0.3cm}
\begin{subfigure}[t]{0.5\textwidth}
  \centering
  \includegraphics[scale=0.32]{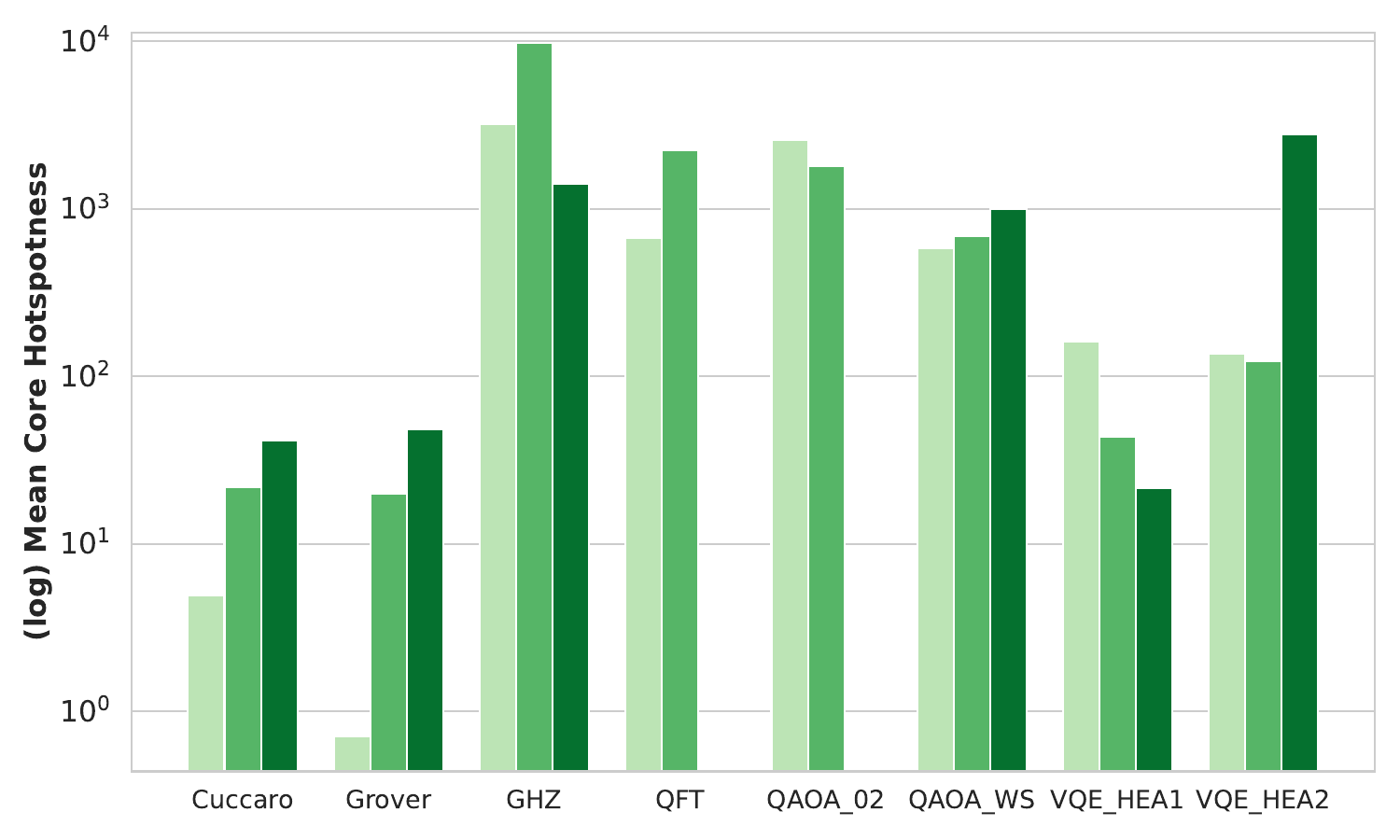}
\end{subfigure}%
\begin{subfigure}[t]{0.5\textwidth}
  \centering
  \includegraphics[scale=0.32]{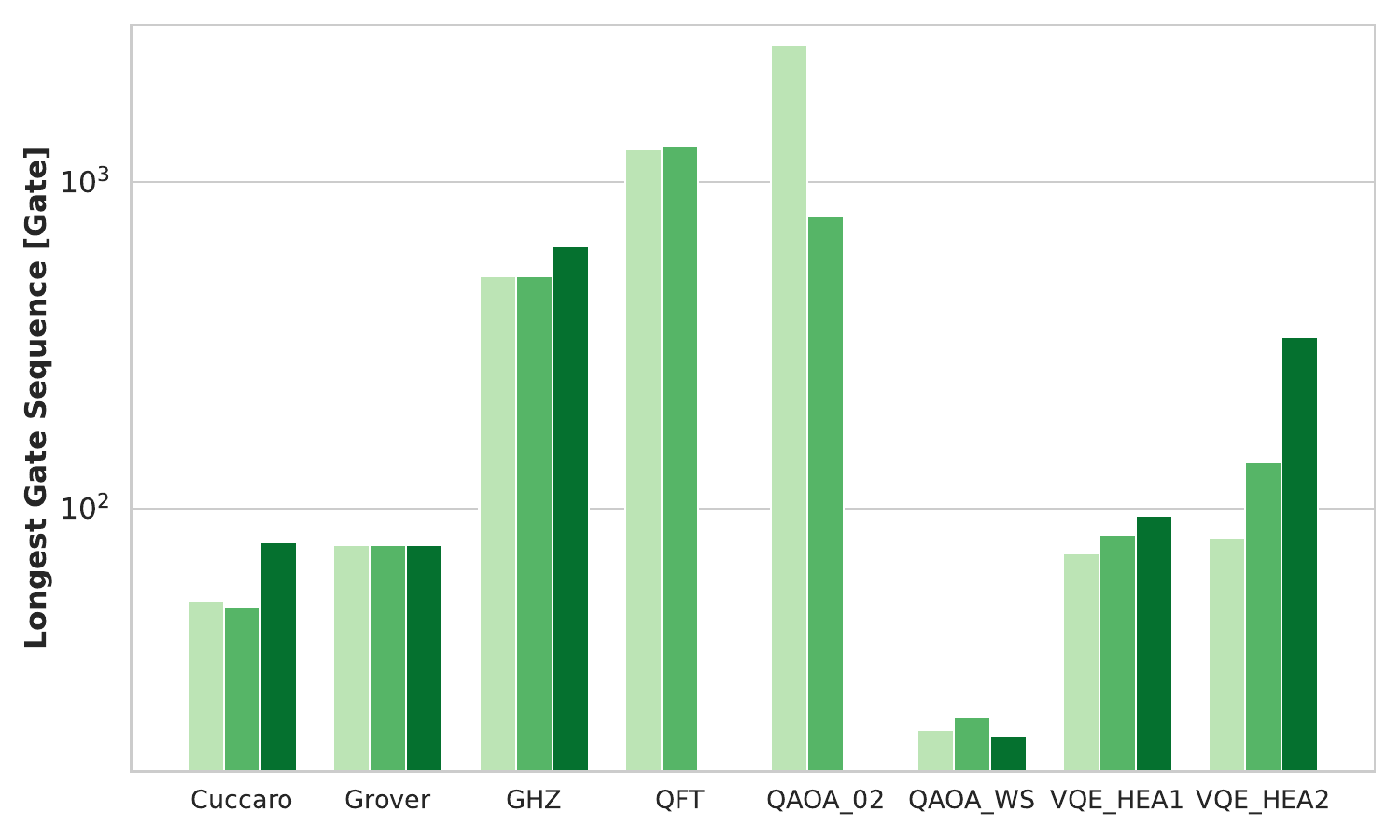}
\end{subfigure}%
\vspace{0.3cm}
\begin{subfigure}[t]{0.5\textwidth}
  \centering
  \includegraphics[scale=0.32]{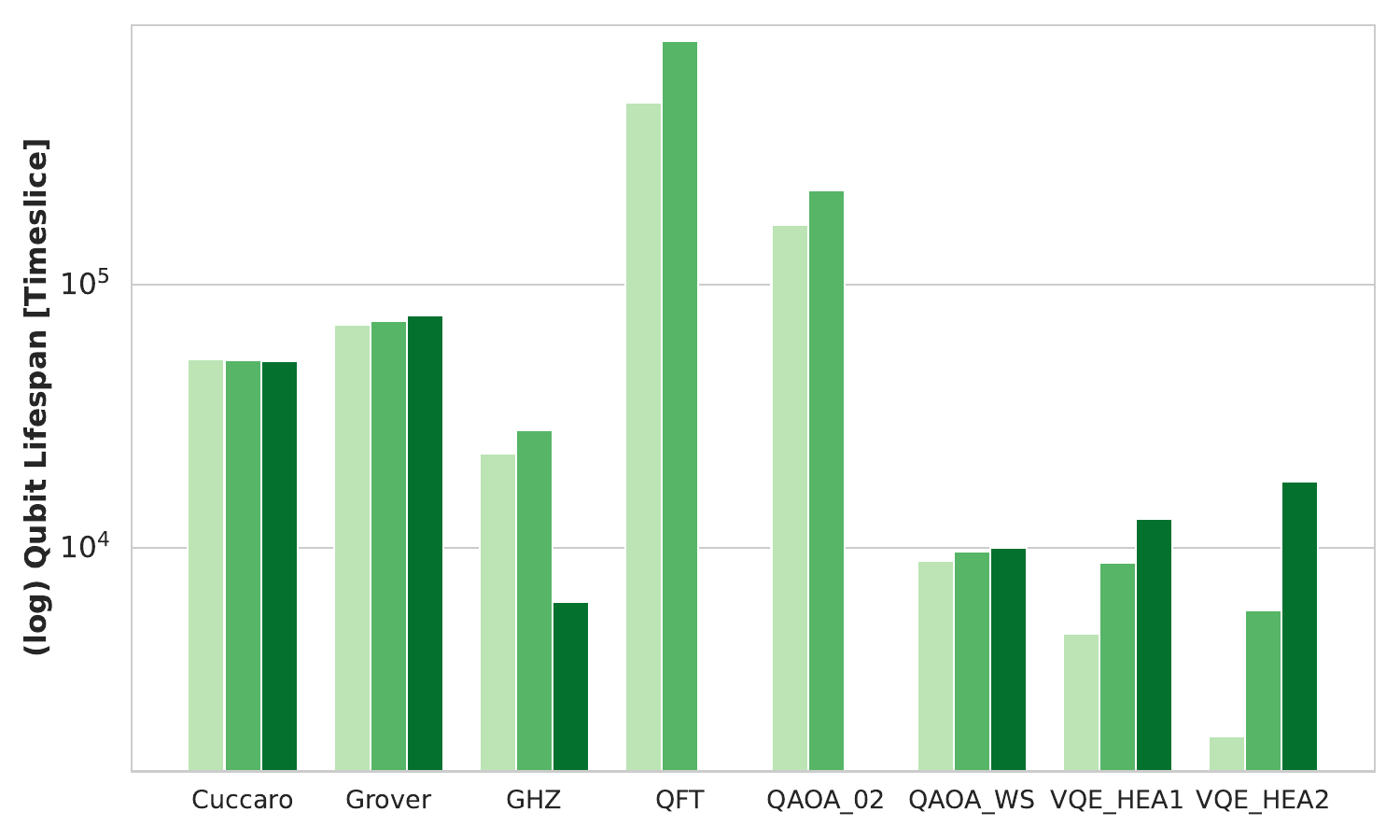}
\end{subfigure}%
\begin{subfigure}[t]{0.5\textwidth}
  \centering
  \includegraphics[scale=0.32]{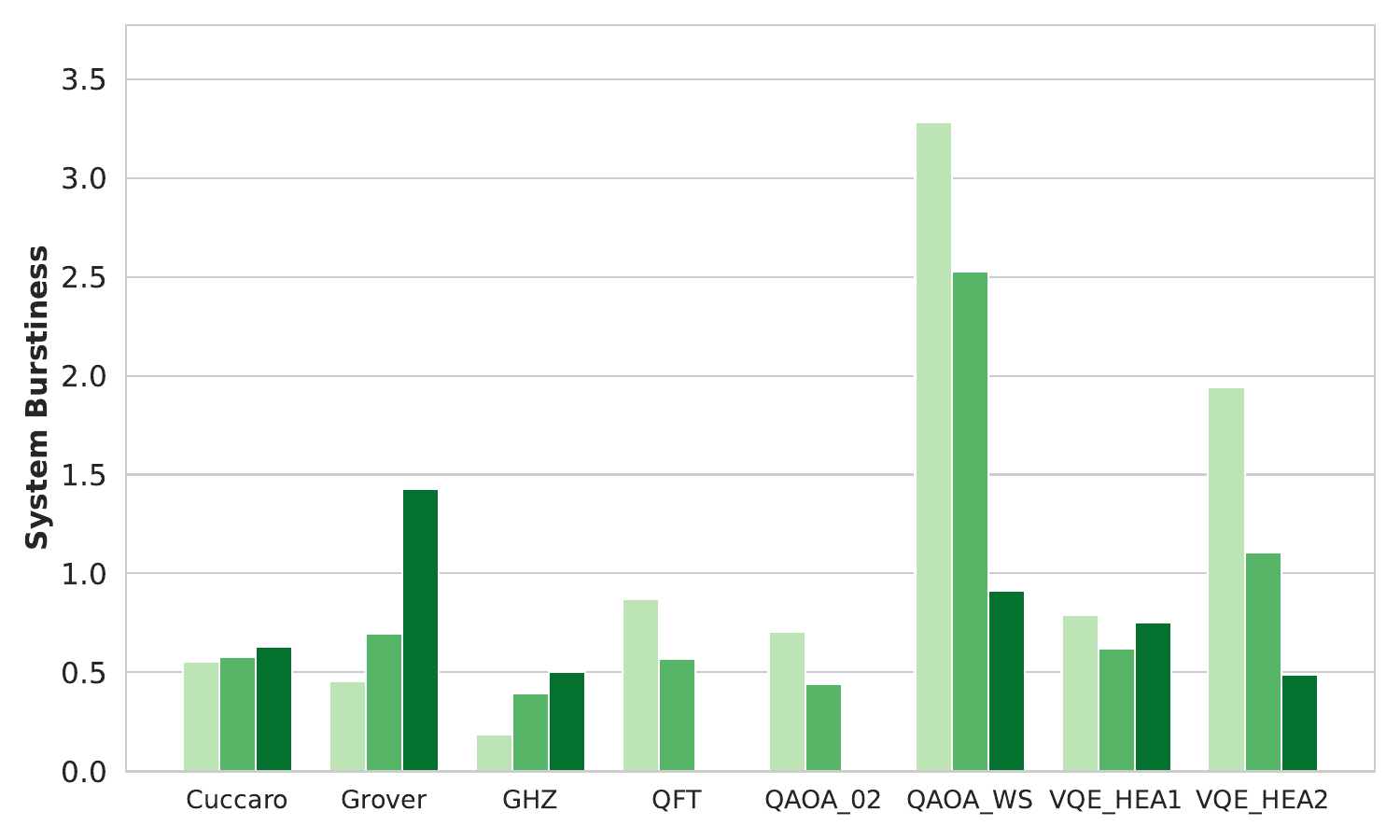}
\end{subfigure}%
\vspace{0.3cm}
\begin{subfigure}[t]{0.5\textwidth}
  \centering
  \includegraphics[scale=0.32]{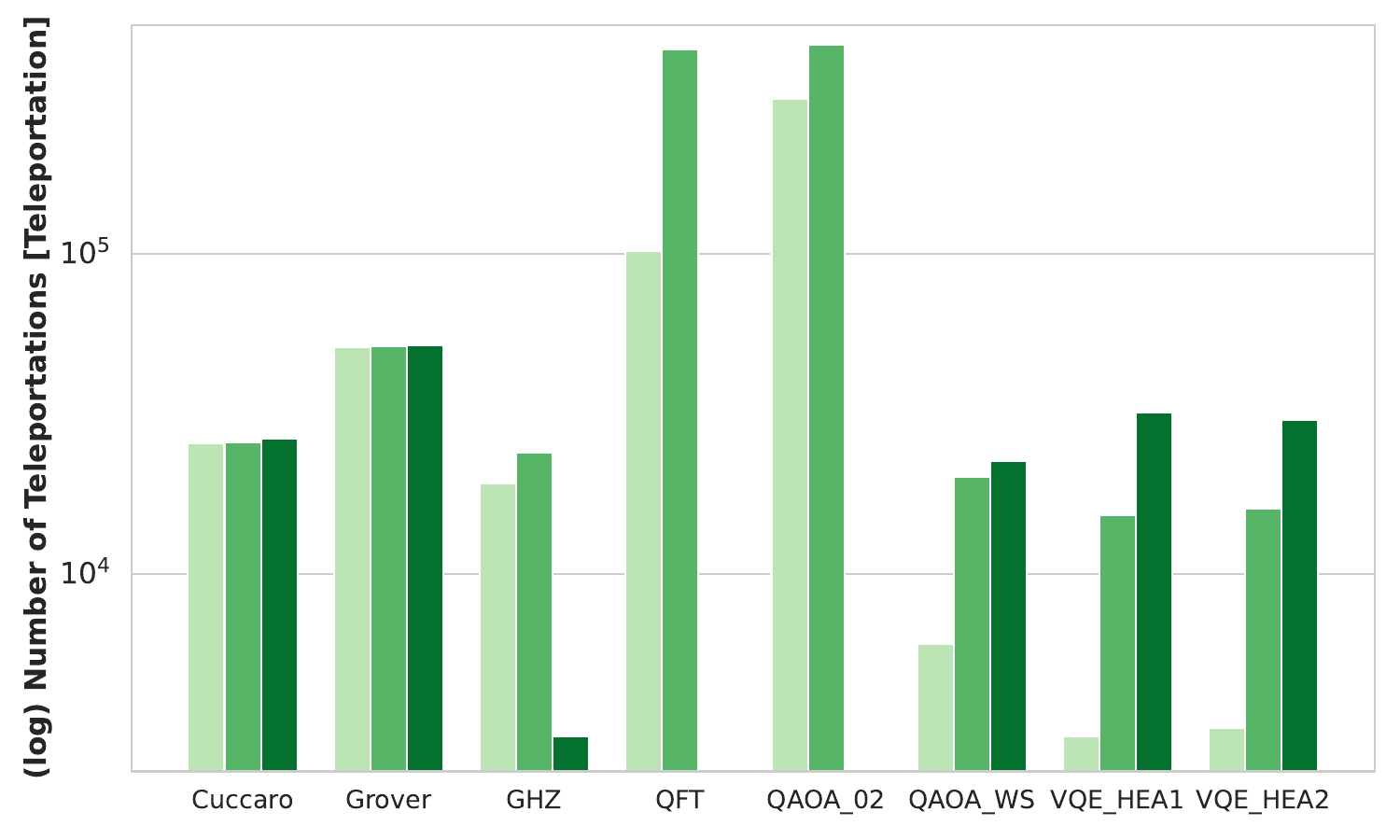}
\end{subfigure}%
\begin{subfigure}[t]{0.5\textwidth}
  \centering
  \includegraphics[scale=0.32]{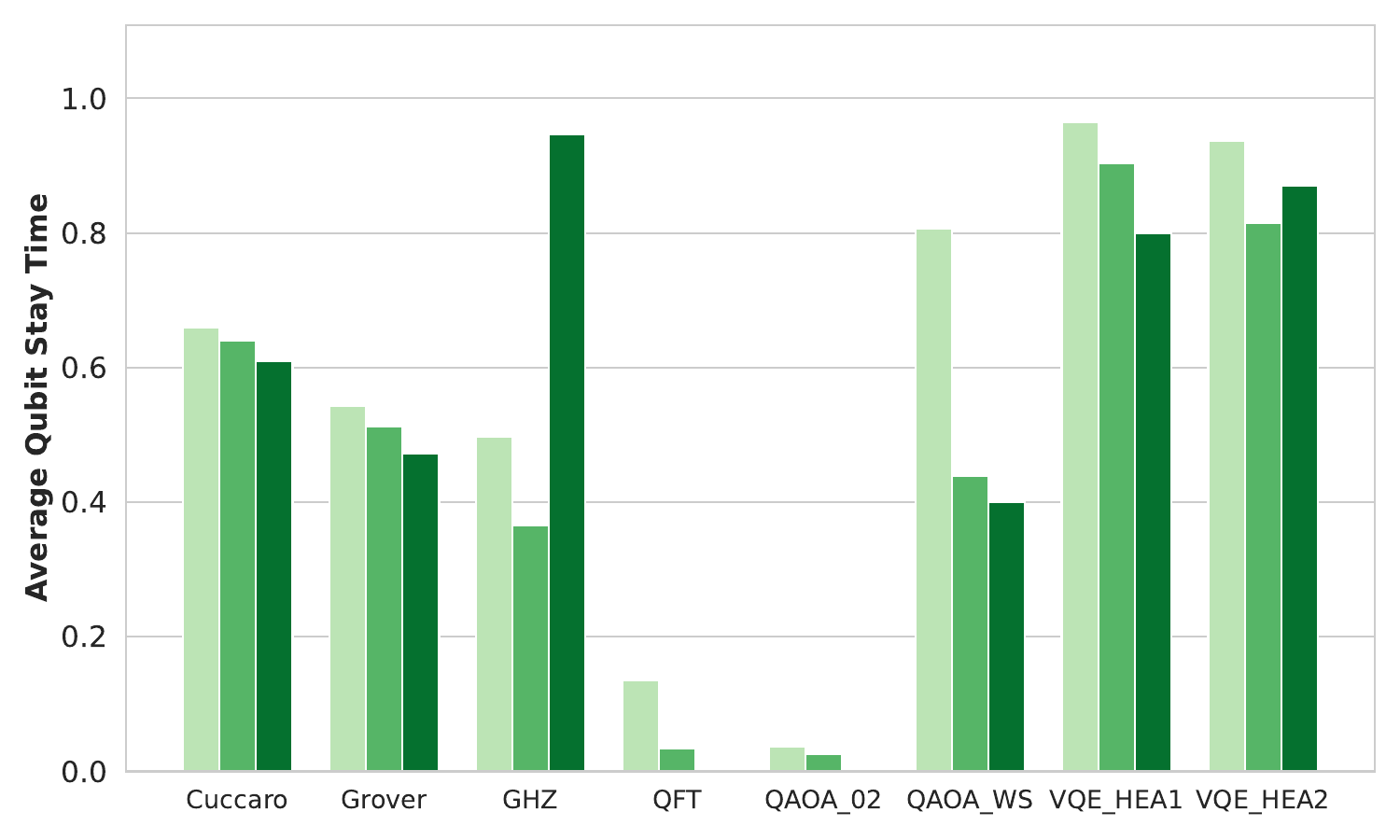}
\end{subfigure}%
\vspace{0.3cm}
\caption{Performance metrics for the selected algorithms executed on modular architectures of 4, 16, and 32 cores, respectively, supporting 512 qubits in the weak scaling.}
  \label{weak_scaling}
\end{figure*}

We analyze the performance of the system in the strong scaling to evaluate the communication overhead as we add more cores, and its impact on the computation process. We fix the size of a single core to 16 qubits. We aimed to increase the number of cores and the size of circuits until 1000 qubits. However, due to the high computational cost of certain algorithms and the current limitations of the compiler, it was not possible to execute all circuits at this size. We present the results of each circuit up to the largest size we could successfully implement.

\subsubsection{Workload distribution}
We interpret the distribution of computation and communication operations over the execution timeslices as an indicator of an optimized workload. In principle, communication tasks take longer times than computation operations. Looking at the results in Figure \ref{ccr_strong_scaling} showcasing the distribution of required computational and communication temporal resources over the execution timeslices, we conclude that the circuit logical structure is a deterministic factor in parallelizing the operations. 

Structured circuits that implement sequential two-qubit gates applied on adjacent or localized qubits, mainly the Cuccaro and Grover's, indicate that the computation operations are not delayed by the communications despite the increasing circuit size. The system achieves an optimal rate of parallelizing computation and communication instructions as we increase the number of cores. 

The GHZ circuit exhibits a sequential application of CNOT gates as well. However, since the first qubit is in constant interaction with the rest of the qubits, adding cores results in increasing the likelihood of two-qubit gates applied across cores. The inter-core communications continuously increase at a high cost, hindering the computational operations. As the size of QFT circuit increases, the inter-core communication requirements grow without blocking the computational process. The QAOA\_02 circuit is characterized by random gate application as imposed by the input graph, which explains the scattered pattern of communication and computational tasks, resulting in prohibitive communication execution times. The QAOA\_WS circuit displays a favourable rate of parallel operations, owing to sequential execution of CNOT gates. A similar behaviour is observed for VQE\_HEA\_1 and VQE\_HEA\_2. Essentially, monitoring the temporal resources needed for inter-core communication is crucial, as the communication overhead translates to a higher risk of quantum state loss and delaying computation operations, and hence deteriorating the computational results.

\subsubsection{System performance characterization}
We analyze the computation process and inter-core qubit traffic according to a set of metrics as depicted in Figure \ref{strong_scaling}. 

Quantum circuits vary significantly in the number of gates and the distribution of two-qubit operations, which is reflected in their differing resource requirements. As the circuit size increases, the computation-to-communication ratio approaches 1 for all circuits, indicating that computational tasks dominate the execution process, thereby allocating significant qubit resources to these tasks. Notably, the QAOA\_02 instance shows a lower ratio for an architecture size of 16 cores, due to the higher frequency of communication operations.

The mean core hotspotness depends on the circuit structure and generally increases with circuit size. For instance, the mean core hotspotness of the structured Cuccaro circuit remains low and stable, whereas the communication-intensive QFT circuit exhibits higher rates that escalate rapidly with increasing algorithm size. Similarly, mean qubit hotspotness, longest gate sequence, and qubit lifespan also increase with circuit size, which can be attributed to the increasing circuit depth. The QFT circuit, in particular, showcases a substantial increase when expanding from 4 to 16 cores in terms of mean qubit hotspotness, longest gate sequence, and qubit lifespan, which increase by approximately 2891\%, 392\%, and 1765\%, respectively. For the GHZ state circuit, the longest gate sequence, representing the interaction of the first qubit with all others, surges by 1738\% as the circuit scales from 4 to 60 cores. 

Communication burstiness remains relatively low for all circuits, indicating that the compilation process distributes teleportation operations uniformly across execution timeslices, with the exception of the 60-core VQE\_HEA\_2 instance that exhibits a higher burstiness due to executing numerous parallel CNOT gates. The number of teleportation instructions increases considerably as larger programs are distributed over more smaller-sized cores, particularly for the QFT circuit due to its all-to-one two-qubit interactions. Similarly, the QAOA\_02 circuit shows a high number of teleportations due to its density and scattered pattern of operations. The increase in teleportation operations translates to a low average qubit stay time for QFT and QAOA\_02 at 16 cores, indicating extensive qubit movement to ensure proximity for communication requirements. Conversely, qubits in other circuits remain relatively stationary, as two-qubit gates are mostly applied to adjacent qubits or qubits within the same core.

This analysis demonstrates that while structured circuits exhibit balanced and steady scaling of communication overhead, random circuits and instances with extensive parallel inter-core operations show substantially high and rapid increases in overhead.

\subsection{Weak scaling}

\subsubsection{Workload distribution}
The distribution of computation and communication operations over the execution time in the weak scaling is depicted in Figure \ref{ccr_weak_scaling}. Generally, we note that inter-core communications increases with the addition of more cores; however, the rate of this increase is dependent on the circuit structure. The system achieves a high task parallelization rate particularly for the QAOA\_WS and the VQE circuits. As the GHZ circuit structure requires long-range cross-core operations, the substantial surge in inter-core communication is prohibiting the computational operations. The compiler improves the instruction distribution when the number of cores is large. A similar behaviour is observed in the QAOA\_02 circuit. The low inter-core communication requirements for circuits of a small number of cores is directly related to the high number of qubits placed in each core, favouring localized operations.

\subsubsection{System performance characterization}
We consider the performance metrics to assess the capacity of the system capacity in executing the given algorithms in the weak scaling as presented in Figure \ref{weak_scaling}.

Quantum circuits exhibit varying computation-to-communication ratios across different architecture sizes. Notably, the 4-core GHZ State and 16-core QAOA\_02 circuits display the lowest ratios, suggesting that these circuits have a higher proportion of communication operations relative to computation operations.

We observe a decrease in qubit hotspotness for Grover's, GHZ, and QFT circuits, while other circuits show a continuous increase. In particular, the qubit hotspotness of the GHZ instance decreases by 98\% passing from 4 to 32 cores. The mean core hotspotness generally rises steadily with the number of cores across all circuit types, with a particularly significant increase noted in the VQE\_HEA\_2 instance, and an exceptional decrease for the VQE\_HEA\_1.

The longest gate sequence and qubit lifespan scale proportionally to the circuit size and architecture, with a particularly higher increase rate for circuits exhibiting a high density of cross-core two-qubit gates, such as QFT. This trend highlights the growing complexity and depth of the circuits as they we distribute the circuit over more cores.

Communication burstiness remains relatively low across all circuits, with the exception of QAOA\_WS that shows a bursty traffic specifically for the 4-core instance. The number of teleportations significantly increases across all circuits as programs are distributed over more cores. This rate is particularly pronounced in the QFT and QAOA\_02 circuits, demanding a high number of teleportation instructions as also reported in the strong scaling analysis. Consequently, these circuits exhibit lower average qubit stay times compared to others.

The average qubit stay time is relatively high in the weak scaling regime, as the cores host a large number of qubits, thereby limiting the need for extensive inter-core state exchanges
\section{Discussion} \label{discussion}

In this paper, we investigate the inter-core qubit traffic in a modular quantum system of increasing sizes in the strong scaling and weak scaling. We evaluate the computation process based on circuit compilation of various quantum algorithms within the constraints imposed by inter-core communication. Specifically, we focus on the impact of inter-core communication on the computational process, considering a particular mapping algorithm, an all-to-all qubit connectivity, and assuming the teleportation protocol for inter-core state transfer. 

By analyzing the circuit mapping to physical qubits, we observe the distribution of communication qubits and computation qubits during the program execution. The circuit mapping process efficiency is dependent on both the algorithm structure and the underlying architecture. The algorithm structure determines the number of gates, the two-qubit interactions and the inter-core communication requirements. It would be pertinent to investigate the adequate trade-off of the number of qubits per core according to the circuit size and logical structure. Our findings indicate that structured instances exhibit a steady increase in communication overhead. However, circuits with extensive and parallel cross-core operations, in addition to random circuits, may experience significant hindrance of the computational process due to this overhead. Communication qubits should be allocated without penalizing the necessary computation resources, which can be achieved by utilizing the idling qubits in the execution process as communication qubits, provided their quantum states are not used for computation.

In the strong scaling, we note that increasing the size of the problem while keeping a relatively low number of qubits per core, typically amplifies the inter-core communication overhead. More inter-core communications lead to a higher likelihood of qubit movements across cores, resulting in a decreased computational efficiency in terms of runtime. Algorithms that involve short-range two-qubit gates applied sequentially on localized or adjacent qubits demonstrate a better performance in parallelizing communication and computation tasks, as well as efficiently applying the necessary communication tasks without delaying the computation process. On the other hand, circuits with a high number of two-qubit gates applied across chips, even if applied sequentially, indicate a growing amount of inter-core communications that hinder the computation process. Circuits with a high density of parallel two-qubit gates demonstrate a comparable behaviour. 

In the weak scaling, circuits exhibiting short-range, local two-qubit gates applied on adjacent qubits achieve a high parallelization rate of computation and communication tasks. As the number of cores increases, the inter-core communication overhead is expected to grow for circuits with a high density of parallel two-qubit operations. It is noteworthy that the communication overhead on the computation process is less prevalent in the weak scaling since the number of qubits per core is higher, and therefore less inter-core state exchanges are required. 

While the all-to-all connectivity is practically unattainable as contemporary quantum processors are constrained by limited processor topologies, the all-to-all connectivity model represents a theoretical upper limit of system performance in terms of inter-core communication. This serves as a benchmark against which the performance of more realistic, constrained architectures can be evaluated. Additionally, in this simplifying assumption for the compilation process, it is easier to conceptualize quantum algorithms as it eliminates the need to consider complex routing and swapping operations, which optimization is an active area of research.

Using the proposed performance metrics, we introduce a framework for estimating the inter-core communication workload and benchmarking modular quantum computers. When considering more realistic, restricted processor topologies, we anticipate the emergence of significantly higher workloads due to the limitations in intra- and inter-core connectivity. By setting a theoretical benchmark as presented in this work, we provide a target workload as a guide toward achieving optimal networking characteristics in practical quantum computing architectures.

Incorporating parameters such as the estimated duration of teleportation operations across cores, realistic processor topologies, and the state transfer process fidelity would enable a quantitative assessment of the impact of communication overhead on the computational process. This analysis would assist hardware designers in determining the optimal number of qubits per core to build robust modular architectures with reliable interconnects that enable parallel operations, and network designers in identifying ideal communication channel parameters that minimize state transfer fidelity losses and latencies. Additionally, the presented results suggest that designing algorithms employing short-range, local two-qubit gates applied on adjacent qubits, or qubits placed within the same processor, is better adapted to the execution on modular architectures. While inter-core communication are the key feature of modular processors, reducing the communication overhead is crucial for a more reliable computational process.
\section{Conclusions and Outlook} \label{conclusion}

As the modular quantum computing technology is developing, we present a qualitative characterization of the inter-core qubit traffic in multicore architectures supporting around 1000 qubits, representing the computation and inter-core communication workloads in large-scale architectures. 

In this work, we demonstrate that circuit mapping is a crucial step in the compilation process that considers essentially the circuit structure and the underlying architecture, and propose a set of performance metrics to assess the communication overhead and computation resources necessary to execute quantum application algorithms on large-scale modular processors. While the necessary computational and communication resources highly depend on the input circuit's structure, we show that all the executed quantum circuits are parallelizeable with significant resources allocated to the computational tasks, the resource usage tends to increase as we scale the circuit size, the temporal locality increases proportionally to the number of cores, and the spatial locality increases respectively to the number of qubits per core.

Our work contributes to setting the foundations of benchmarking modular quantum processors by proposing performance metrics and an application-oriented characterization of modular quantum computers. In future works, we will explore the performance of the system considering the circuit structure and the topology of quantum processors, seeking to identify the ideal hardware architecture for specific categories of circuits from a co-design perspective.

\printbibliography

\begin{appendices}
\section*{Appendix}
\section{Implemented quantum circuits} \label{implemeted-circuits}
\begin{figure*}[h]
    \centering 
    
    \begin{subfigure}[t]{0.45\textwidth}
        \centering 
        \includegraphics[width=\textwidth]{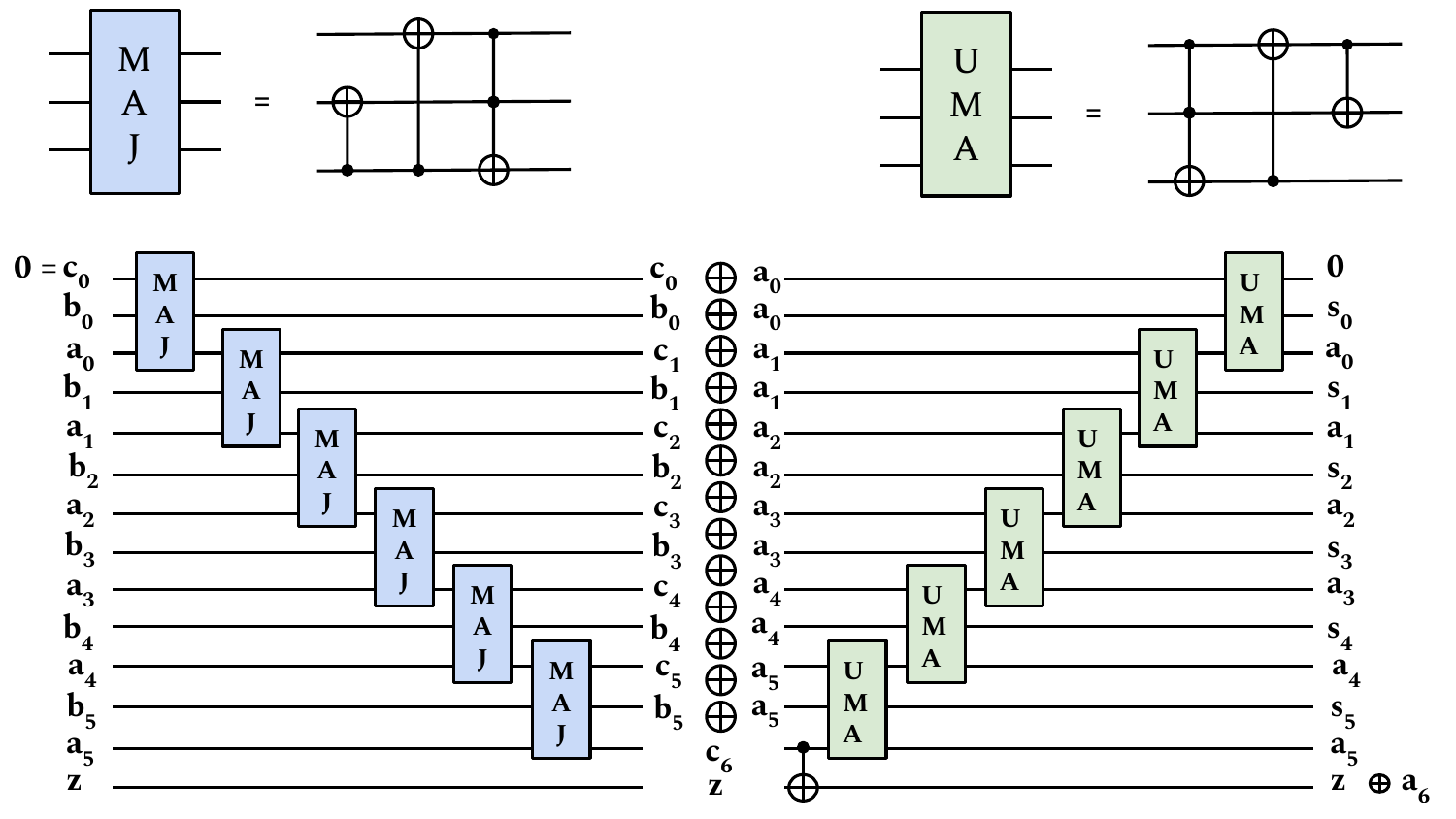} 
        \caption{Ripple-carry Cuccaro Adder}
    \end{subfigure}%
    \hfill 
    \begin{subfigure}[t]{0.45\textwidth}
        \centering
        \includegraphics[width=\textwidth]{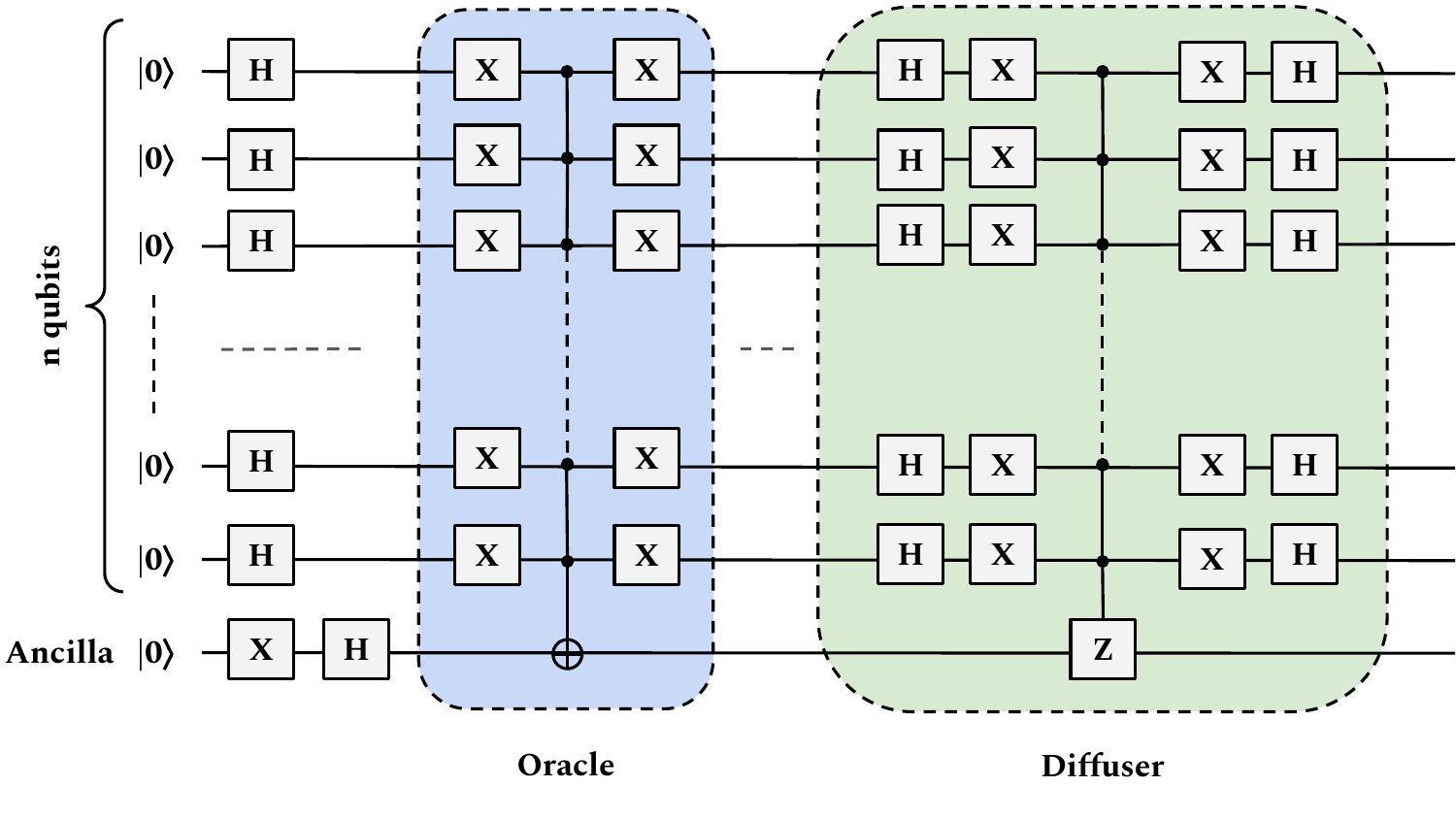}
        \caption{Grover's Search main routine}
    \end{subfigure}
    
    \vspace{0.1cm}
    
    \begin{subfigure}[t]{0.45\textwidth}
        \centering
        \includegraphics[width=\textwidth]{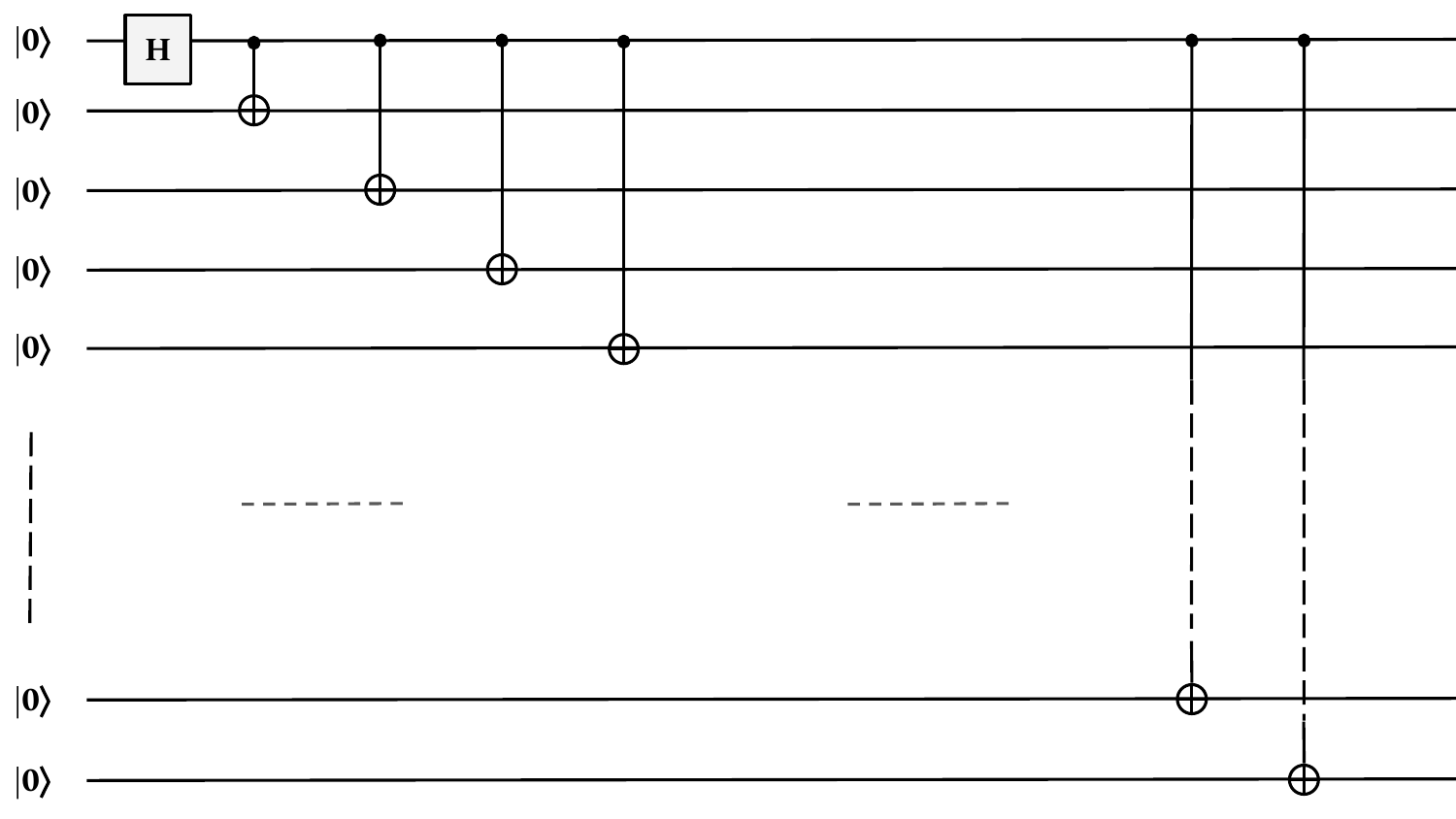}
        \caption{GHZ State}
        \label{ghz}
    \end{subfigure}%
    \hfill
    \begin{subfigure}[t]{0.45\textwidth}
        \centering
        \includegraphics[width=\textwidth]{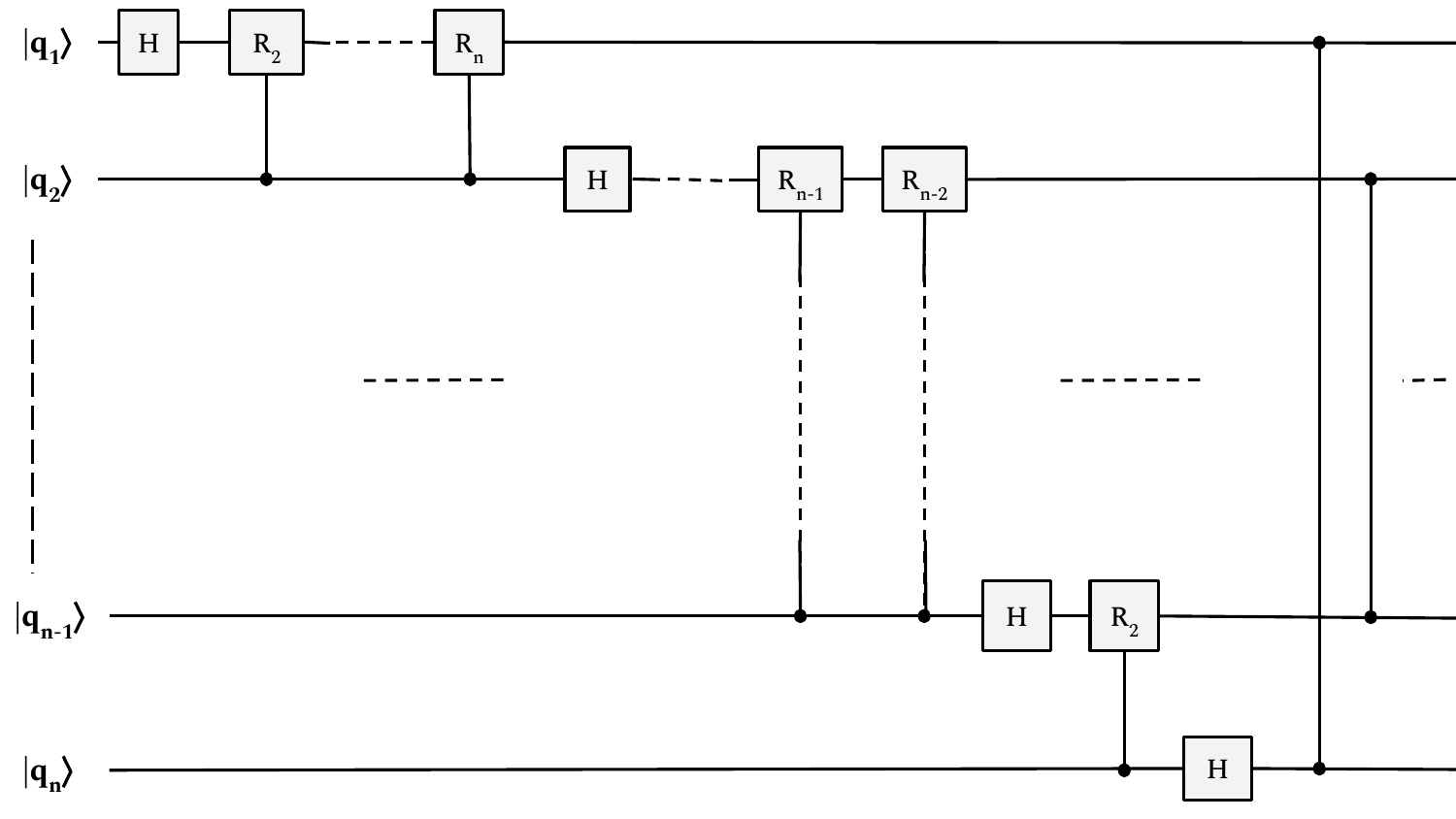}
        \caption{Quantum Fourier Transform}
    \end{subfigure}
    
    \vspace{0.1cm}
    
    \begin{subfigure}[t]{0.45\textwidth}
        \centering
        \includegraphics[width=\textwidth]{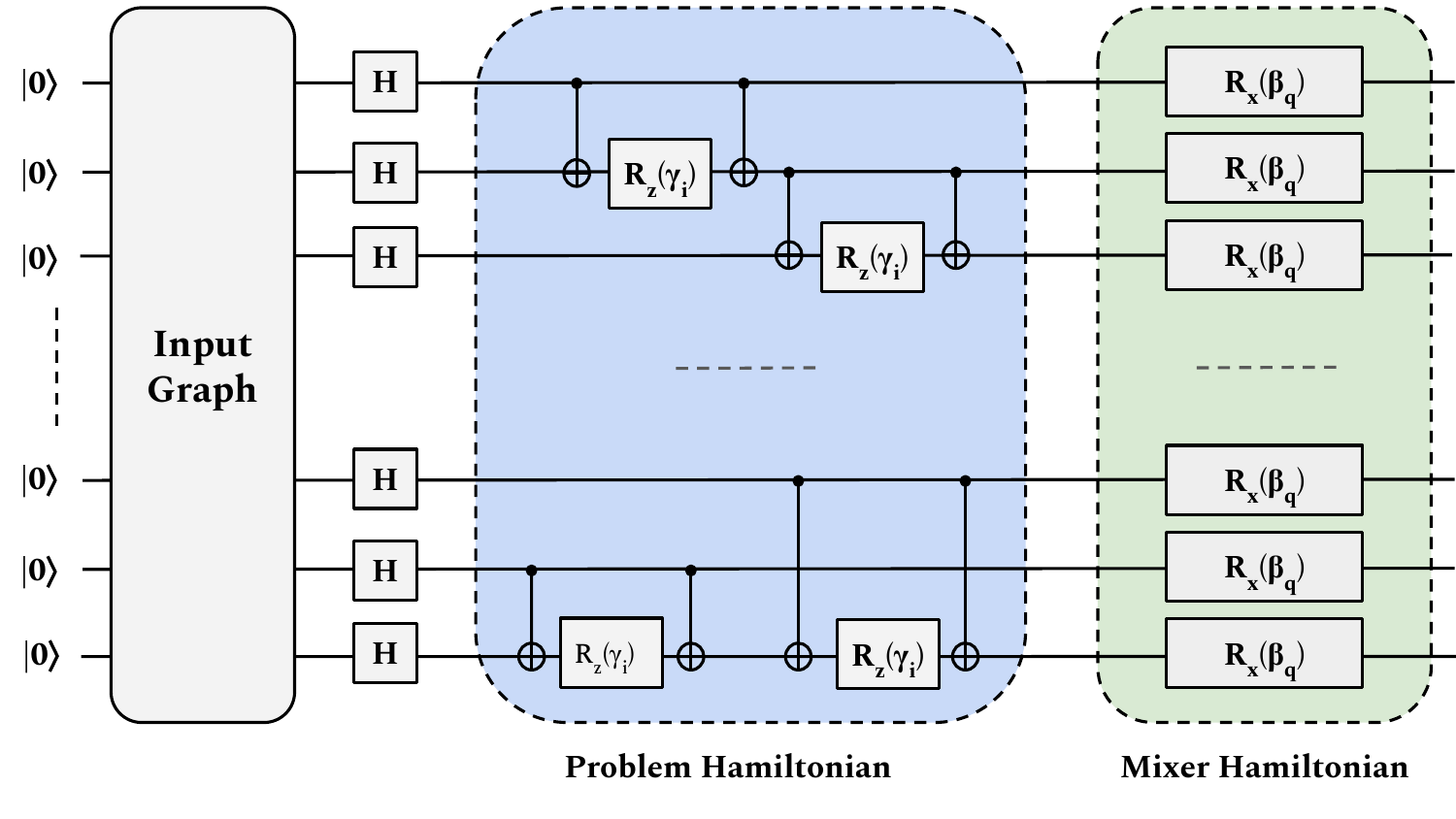}
        \caption{QAOA circuit}
    \end{subfigure}%
    \hfill
    \begin{subfigure}[t]{0.45\textwidth}
        \centering
        \includegraphics[width=\textwidth]{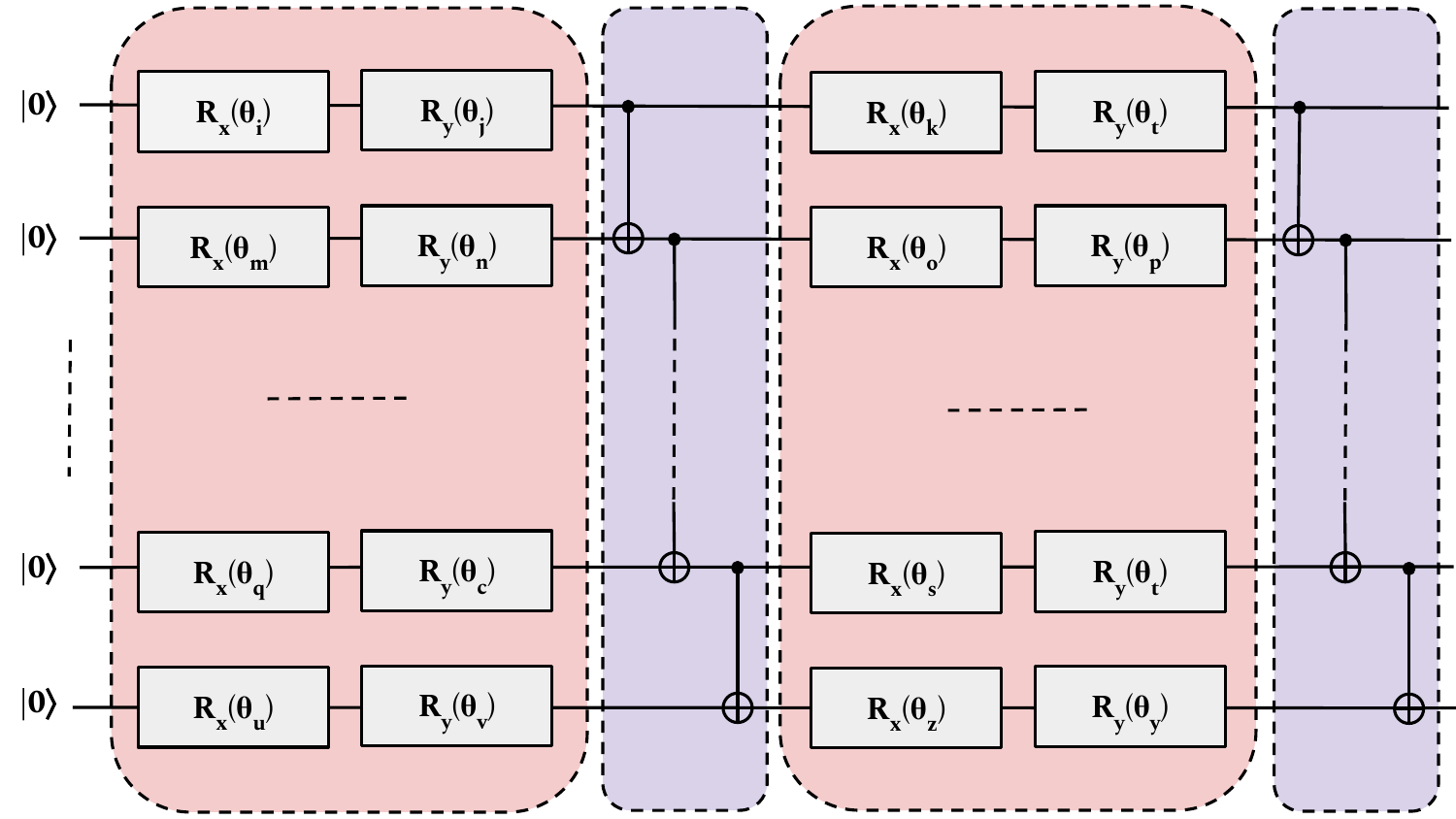}
        \caption{VQE Hardware-efficient ansatz 1}
    \end{subfigure}
    
    \vspace{0.1cm}
    
    \begin{subfigure}[t]{0.45\textwidth}
        \centering
        \includegraphics[width=\textwidth]{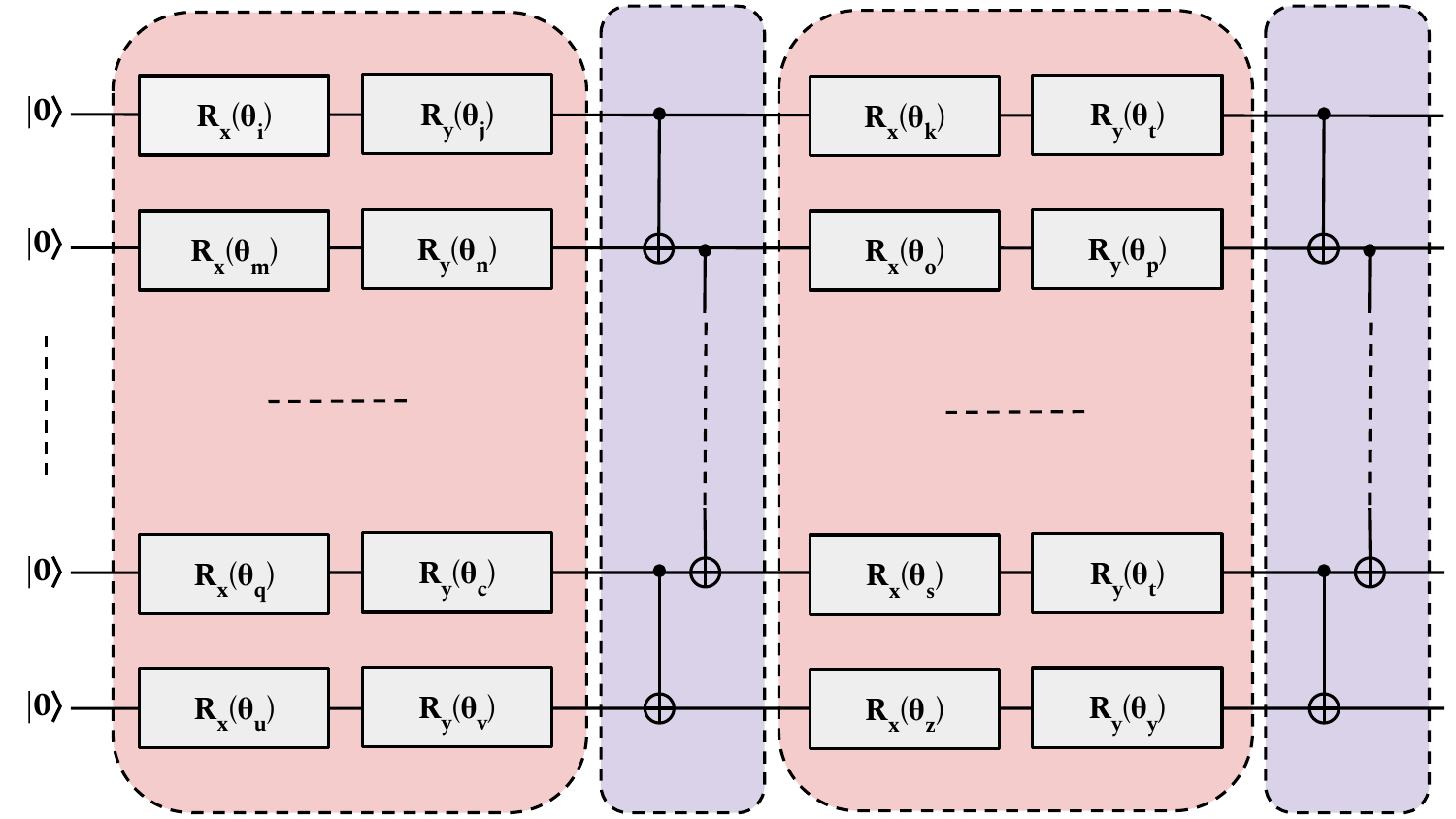} 
        \caption{VQE Hardware-efficient ansatz performing single-qubit x- and y-rotations, and nearest-neighbor two-qubit CNOT gates applied in parallel}
    \end{subfigure}
    
    \caption{Implemented quantum circuits}
    \label{circuits}
\end{figure*}

We showcase the graphic representation of the implemented circuits in Figure \ref{circuits}, which present the logical structure and the gate distribution in the circuits.

\section{Estimated gate count} \label{gate-count}

\begin{enumerate}
    \item \textbf{The ripple-carry Cuccaro adder:} The gate count for this circuit depends on the number of qubits $N$ representing the binary numbers being added. The gate count can be approximated as the number of Hadamard gates for state preparation $\approx O(N)$, and the two-qubit gates applied on the $N-1$ carry qubits estimated as $\approx O(N)$. Therefore, the total gate count is $\approx O(N)$.

    \item \textbf{The Grover's search main routine:} The number of iterations, denoted $k$, applied to find the target element with a high probability is proportional to the total number of elements in the database. The gate count in a single iteration can be approximated as the number of single-qubit gates (Hadamard and X) used for state preparation $\approx O(N)$, the number of single-qubit gates $\approx O(N)$ and two-qubit gates applied to pairs of qubits $\approx O(N)$ in the oracle operator, the number of single-qubit gates $\approx O(N)$ and two-qubit gates applied to pairs of qubits $\approx O(N)$ in the diffuser operator. The total gate count can then be approximated as $\approx O(k \times N)$.

    \item \textbf{The GHZ State circuit:} The gate count in this circuit is estimated as the Hadamard gate applied to the first qubit and the CNOT gates applied to $N-1$ qubits. The total gate count is then estimated as $\approx O(N)$.

    \item \textbf{The QFT circuit:} The gate count in this circuit is estimated as the number of Hadamard gates applied to each qubit $= O(N)$, and the number of Controlled-Phase gates $\approx O((N-1) \times (N-q)) \approx O(N^{2})$, and the number of SWAP operations applied to pairs of qubits $\approx O(N)$. The total gate count is then approximated as $\approx O(N^{2})$.

    \item \textbf{The QAOA MaxCut ansatz circuit:} The number of ansatz layers, denoted $l$, is a tunale parameter. The gate count in a single layer is approximated as the number of Hadamard gates required for state preparation $\approx O(N)$, the number of single-qubit z-rotation gates $\approx O(N)$ and two-qubit gates applied to pairs of qubits $\approx O(N)$ in the Problem Hamiltonian operator, and the number of single-qubit x-rotation gates $\approx O(N)$ applied in the Mixer Hamiltonian operator to all qubits. The total gate count is then estimated as $\approx O(l \times N)$

    \item \textbf{The VQE Hardware Efficient ansatz circuit:} The number of ansatz layers, denoted $l$, is a tunale parameter. The gate count in a single layer is approximated as the number of single-qubit x-rotation and y-rotation $\approx O(N)$ applied to qubits, and the number of CNOT gates applied to pairs of qubits $\approx O(N)$. The total gate count is then estimated as $\approx O(l \times N)$
\end{enumerate}

\section{Circuit mapping traces} \label{traces}
\begin{figure*}[h]
    \centering
    \begin{subfigure}[t]{0.37\textwidth}
        \centering
        \captionsetup{justification=raggedright,singlelinecheck=false}
        \includegraphics[scale=0.29]{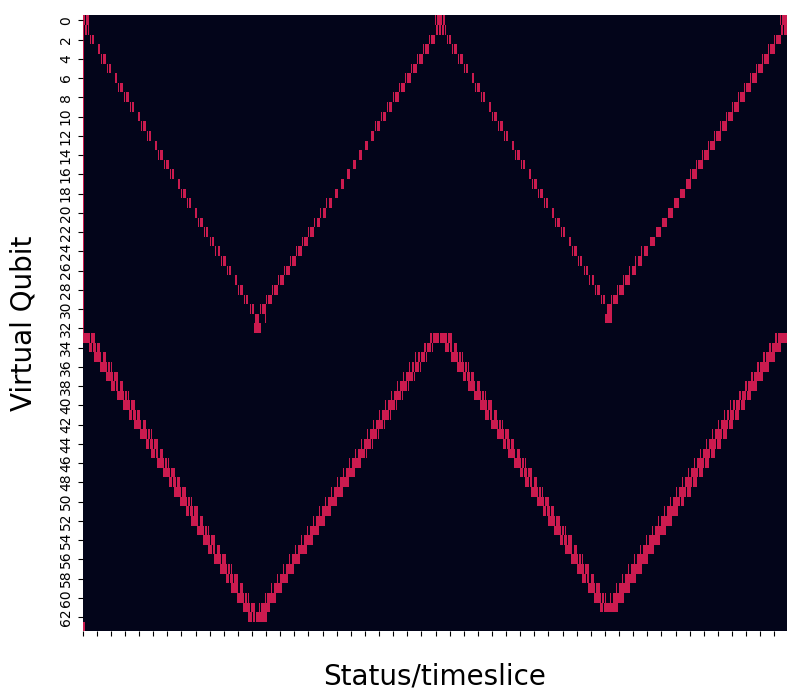}
        \caption{Grover's Search main routine}
    \end{subfigure}%
    \begin{subfigure}[t]{0.37\textwidth}
        \centering
        \captionsetup{justification=raggedright,singlelinecheck=false}
        \includegraphics[scale=0.29]{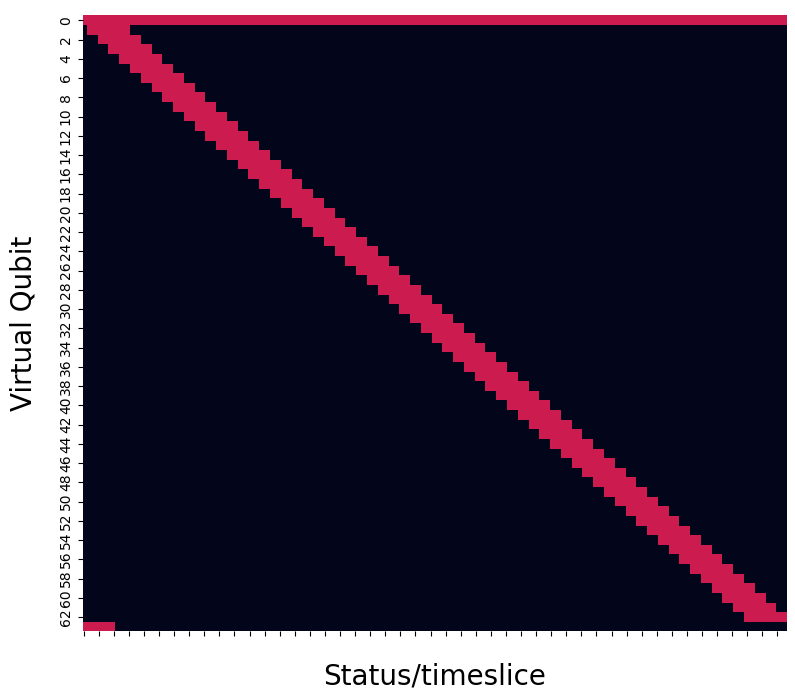}
        \caption{GHZ State}
    \end{subfigure}

    \begin{subfigure}[t]{0.37\textwidth}
        \centering
        \captionsetup{justification=raggedright,singlelinecheck=false}
        \includegraphics[scale=0.29]{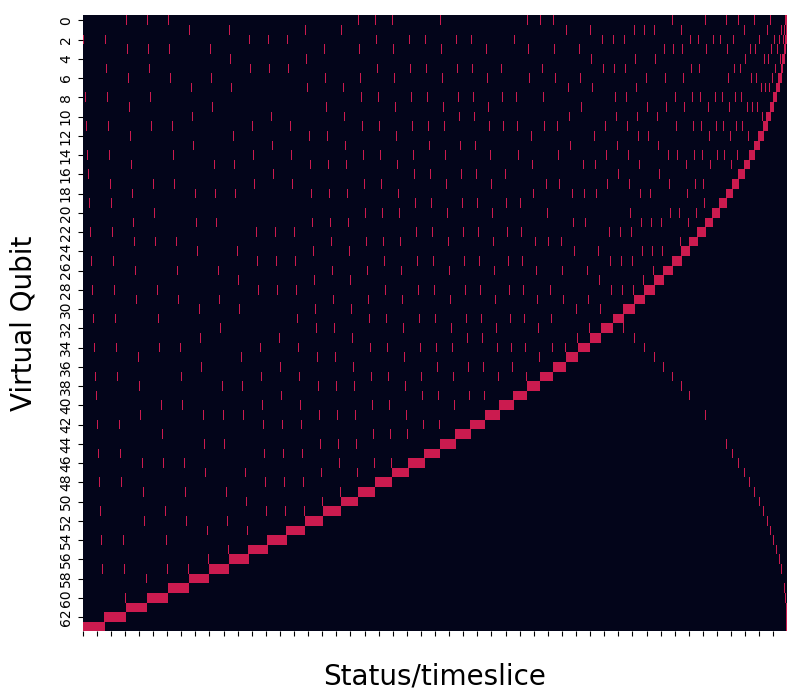}
        \caption{Quantum Fourier Transform}
    \end{subfigure}%
    \begin{subfigure}[t]{0.37\textwidth}
        \centering
        \captionsetup{justification=raggedright,singlelinecheck=false}
        \includegraphics[scale=0.29]{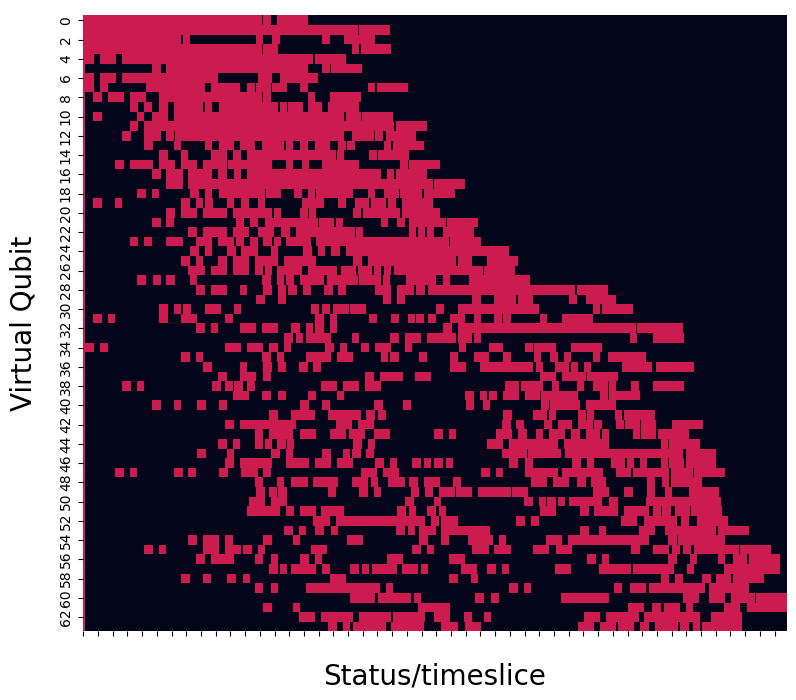}
        \caption{QAOA\_02}
    \end{subfigure}

    \begin{subfigure}[t]{0.37\textwidth}
        \centering
        \captionsetup{justification=raggedright,singlelinecheck=false}
        \includegraphics[scale=0.29]{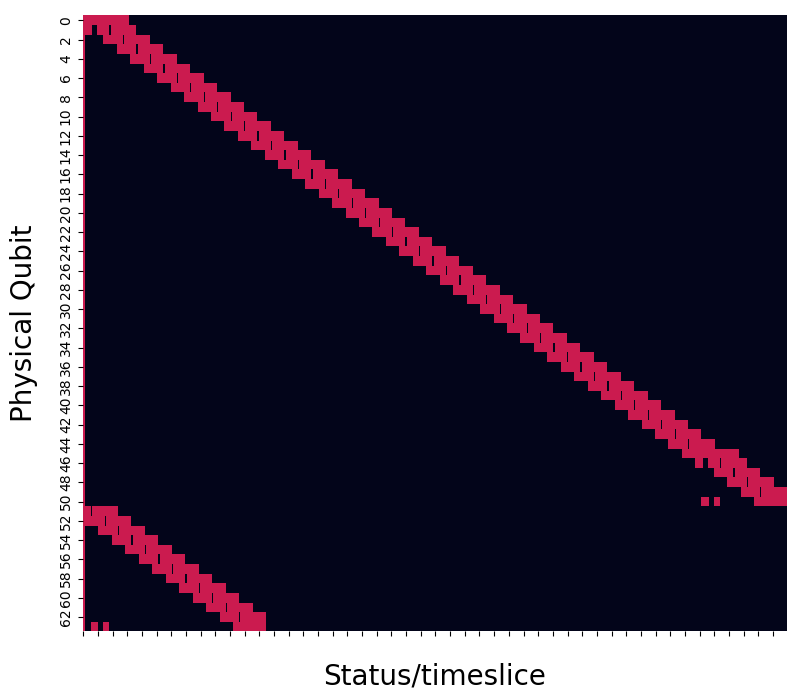}
        \caption{QAOA\_WS}
    \end{subfigure}%
    \begin{subfigure}[t]{0.37\textwidth}
        \centering
        \captionsetup{justification=raggedright,singlelinecheck=false}
        \includegraphics[scale=0.29]{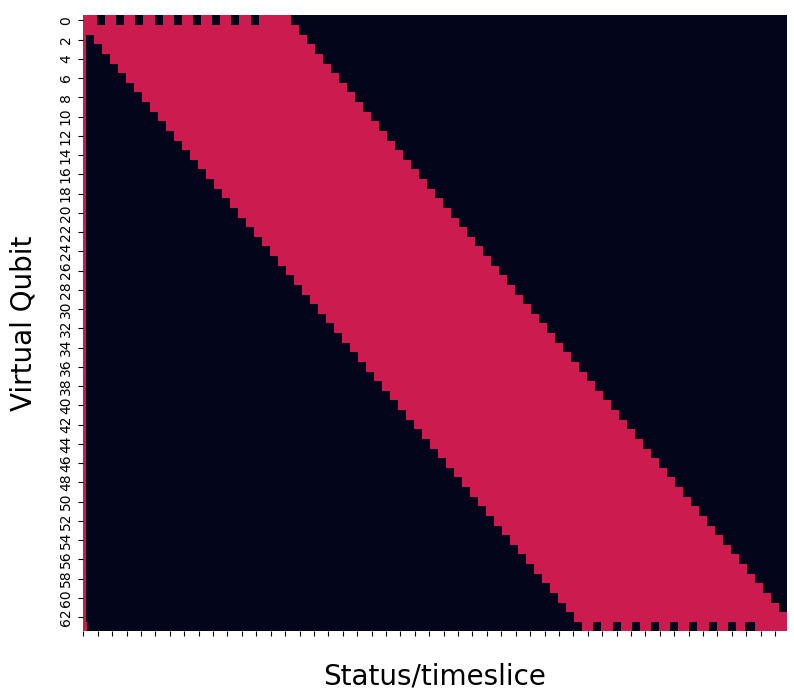}
        \caption{VQE HEA\_1}
    \end{subfigure}

    \begin{subfigure}[t]{0.37\textwidth}
        \centering
        \captionsetup{justification=raggedright,singlelinecheck=false}
        \includegraphics[scale=0.29]{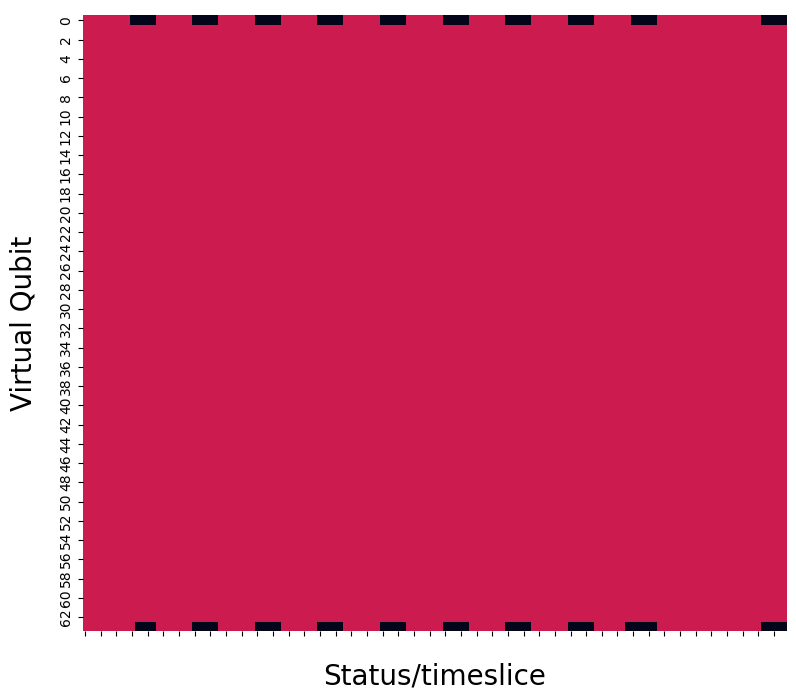}
        \caption{VQE HEA\_2}
    \end{subfigure}
    
    \caption{Logical structure of selected circuits of 64 qubits. Computation times are represented in red.}
    \label{virt_circuits}
\end{figure*}

\begin{figure*}[h]
    \centering
    \begin{subfigure}[t]{0.37\textwidth}
        \centering
        \captionsetup{justification=raggedright,singlelinecheck=false}
        \includegraphics[scale=0.29]{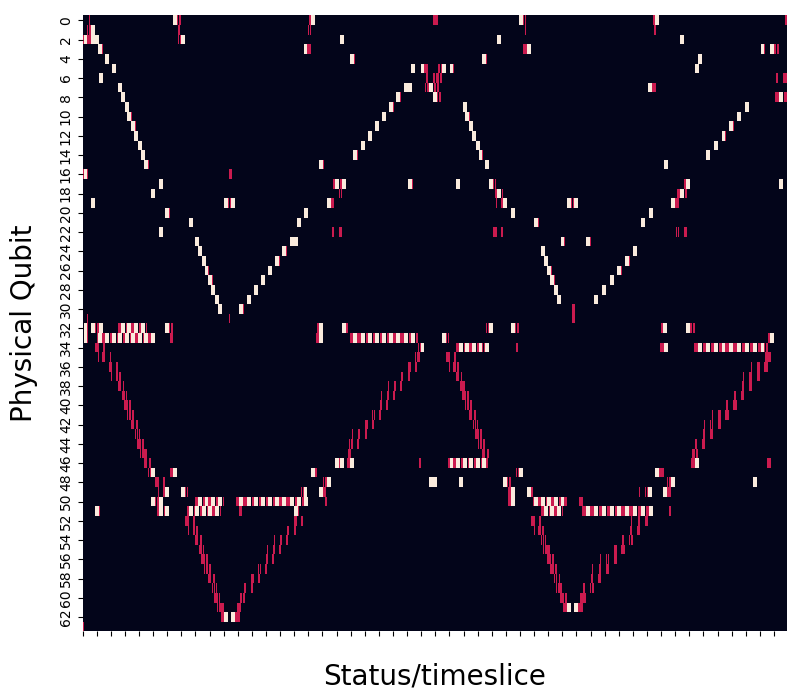}
        \caption{Grover's Search main routine}
    \end{subfigure}%
    \begin{subfigure}[t]{0.37\textwidth}
        \centering
        \captionsetup{justification=raggedright,singlelinecheck=false}
        \includegraphics[scale=0.29]{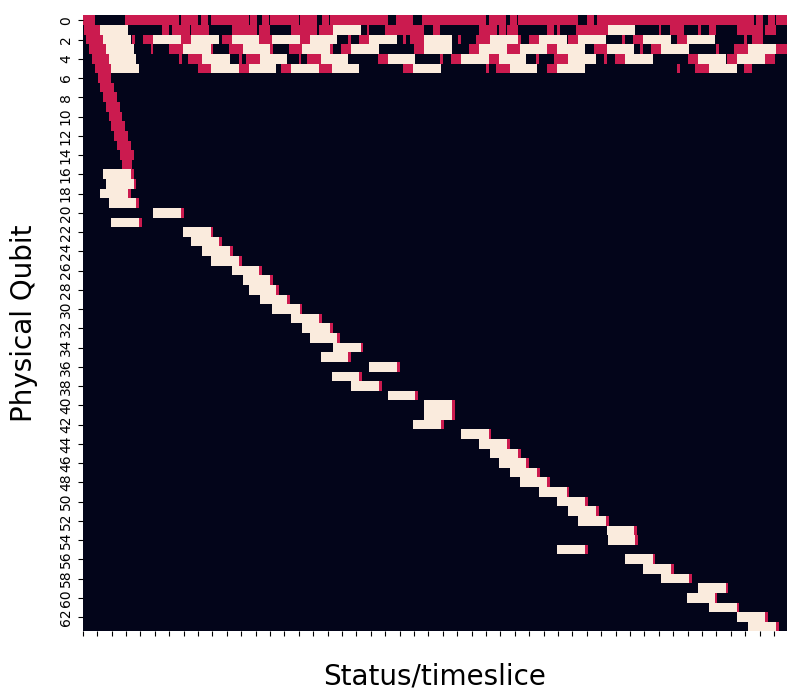}
        \caption{GHZ State}
    \end{subfigure}

    \begin{subfigure}[t]{0.37\textwidth}
        \centering
        \captionsetup{justification=raggedright,singlelinecheck=false}
        \includegraphics[scale=0.29]{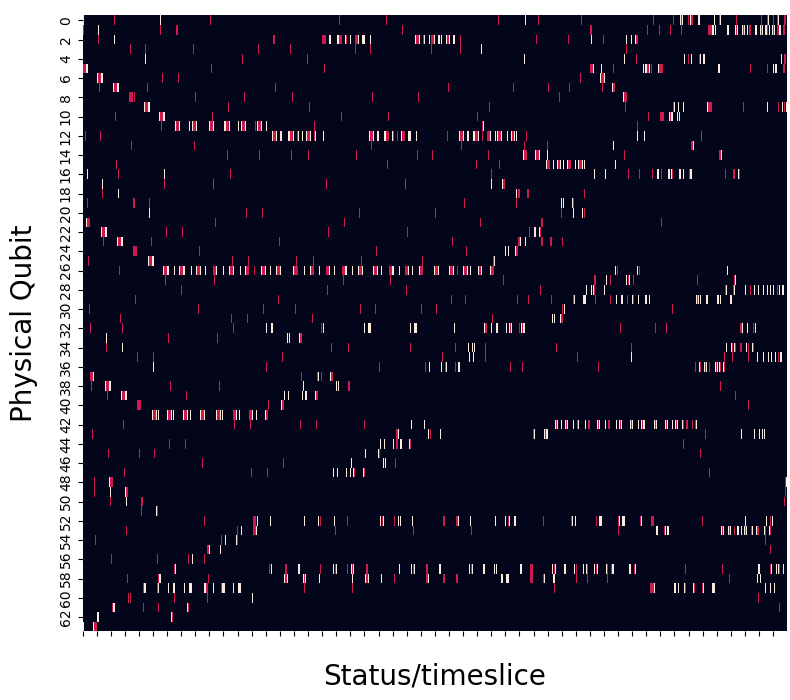}
        \caption{Quantum Fourier Transform}
    \end{subfigure}%
    \begin{subfigure}[t]{0.37\textwidth}
        \centering
        \captionsetup{justification=raggedright,singlelinecheck=false}
        \includegraphics[scale=0.29]{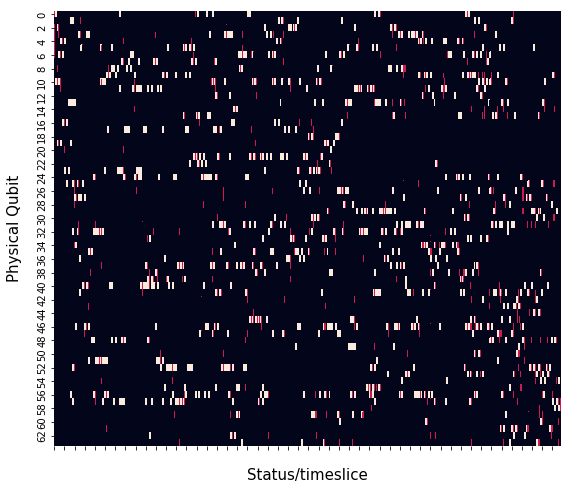}
        \caption{QAOA\_02}
    \end{subfigure}

    \begin{subfigure}[t]{0.37\textwidth}
        \centering
        \captionsetup{justification=raggedright,singlelinecheck=false}
        \includegraphics[scale=0.29]{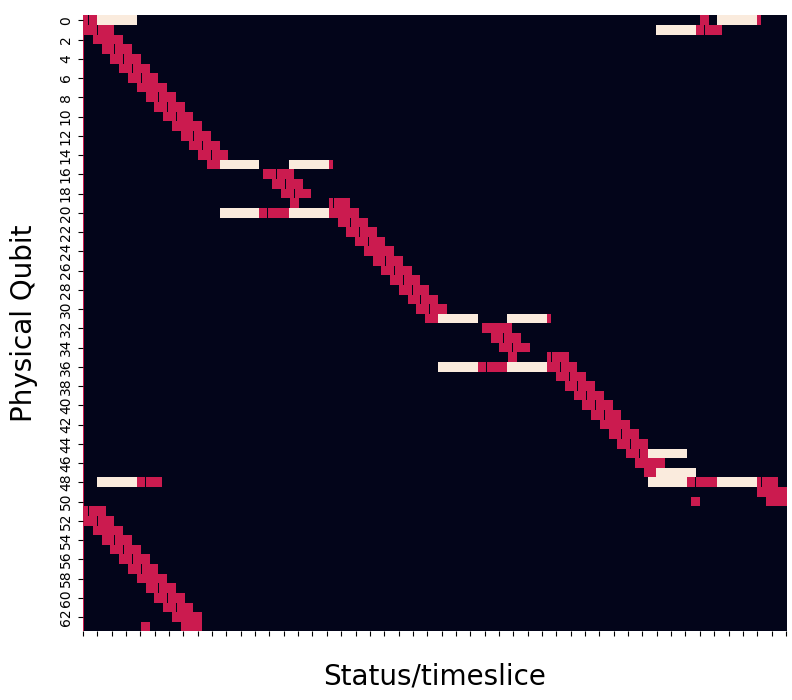}
        \caption{QAOA\_WS}
    \end{subfigure}%
    \begin{subfigure}[t]{0.37\textwidth}
        \centering
        \captionsetup{justification=raggedright,singlelinecheck=false}
        \includegraphics[scale=0.29]{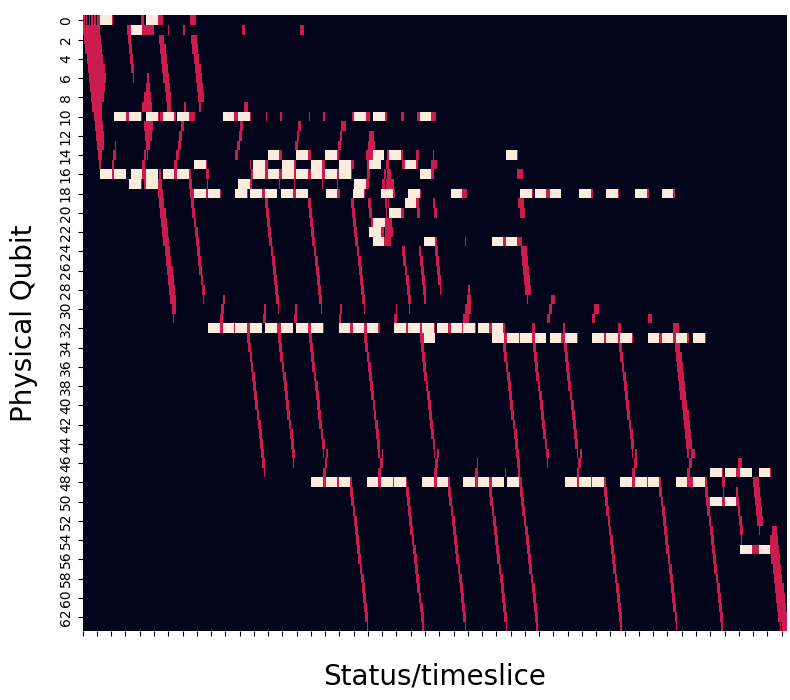}
        \caption{VQE HEA\_1}
    \end{subfigure}

    \begin{subfigure}[t]{0.37\textwidth}
        \centering
        \captionsetup{justification=raggedright,singlelinecheck=false}
        \includegraphics[scale=0.29]{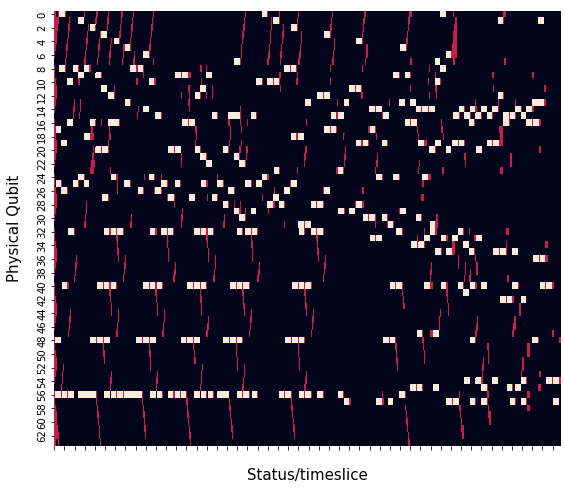}
        \caption{VQE HEA\_2}
    \end{subfigure}
    
    \caption{Circuit mapping traces to physical qubits on a modular architecture of 16 qubits $\times$ 4 cores with all-to-all connectivity. Computation, communication, and idling times are represented in red, white, and black, respectively}
      \label{phys_circuits}
\end{figure*}

We hereby present the the circuit traces for logical structure and physical mapping of the selected algorithms in Figures \ref{virt_circuits} and \ref{phys_circuits}, respectively. We observe the initial circuit structures: the Grover's circuit showcasing the iterative oracle-diffuser cycle, the GHZ State circuit exhibiting a sequential CNOT gate ladder applied from the first qubit to all reamining qubits, the QFT instance illustrating the SWAP-ed pattern of a single qubit interaction with all the remaining ones, the random operations in the QAOA\_02 circuit imposed by the random Erdos Renyi input graph as opposed to the short-ranged and highly clustered operations determined by the Watts Strogatz input graph in the QAOA\_WS circuit. The two VQE ansatze implemented demonstrate distinct traces as imposed by order of CNOT gate application. The execution traces of circuit mapping to physical qubits indicate the structural changes made during the compilation process to comply with the the underlying modular architecture and inter-core communication requirements.

\end{appendices}

\end{document}